\documentclass[aip,reprint,cha,amsmath,amsfonts,amssymb]{revtex4-1}
\usepackage{here}
\usepackage{docs}
\usepackage{graphicx}
\usepackage{amsmath,amsfonts,amssymb}
\usepackage{color}
\usepackage{bm}
\usepackage{ascmac} 
\usepackage{threeparttable}
\usepackage[mathscr]{eucal}
\usepackage{titlesec}

\usepackage{picture}
\usepackage{ulem}
\newcommand\redsout{\bgroup\markoverwith{\textcolor{red}{\rule[0.5ex]{2pt}{0.4pt}}}\ULon}

\definecolor{purple}{rgb}{0.8,0.0,0.8}

 \expandafter\ifx\csname package@font\endcsname\relax\else
  \expandafter\expandafter
  \expandafter\usepackage
  \expandafter\expandafter
  \expandafter{\csname package@font\endcsname}%
 \fi

\newcommand{\bra}{\langle}
\newcommand{\ket}{\rangle}
\newcommand{\bre}{\nonumber\\}
\newcommand{\dd}{\partial}
\newcommand{\calF}{{\mathcal F}}

\begin{document}
\title{Bridging single- and multireference domains for electron correlation: spin-extended coupled electron pair approximation }
\author{Takashi Tsuchimochi}
\email{tsuchimochi@gmail.com}
\affiliation{Graduate School of Science, Technology, and Innovation, Kobe University, Kobe, Hyogo 657-0025 Japan}
\author{Seiichiro Ten-no}
\email{tenno@garnet.kobe-u.ac.jp}
\affiliation{Graduate School of Science, Technology, and Innovation, Kobe University, Kobe, Hyogo 657-0025 Japan}
\affiliation{Graduate School of System Informatics, Kobe University, Kobe, Hyogo 657-0025 Japan}
\date{\today}
\begin{abstract}
We propose a size-consistent generalization of the recently developed spin-extended configuration interaction with singles and doubles (ECISD), where a CI wave function is explicitly spin-projected. The size-consistent effect is effectively incorporated by treating quadruples within the formulation of coupled electron pair approximation. As in coupled-cluster theory, quadruple excitations are approximated by a disconnected product of double excitations. Despite its conceptual similarity to the standard single- and multireference analogues, such a generalization requires careful derivation, as the spin-projected CI space is non-orthogonal and overcomplete. Although our methods generally yield better results than ECISD, size-consistency is only approximately retained because the action of a symmetry-projection operator is size-inconsistent. In this work, we focus on simple models where exclusion-principle-violating terms, which eliminate undesired contributions to the correlation effects, are either completely neglected or averaged. These models possess an orbital-invariant energy functional that is to be minimized by diagonalizing an energy-shifted effective Hamiltonian within the singles and doubles manifold. This allows for a straightforward generalization of the ECISD analytical gradients needed to determine molecular properties and geometric optimization.
Given the multireference nature of the spin-projected Hartree--Fock method, the proposed approaches are expected to handle static correlation, unlike single-reference analogues. We critically assess the performance of our methods using dissociation curves of molecules, singlet-triplet splitting gaps, hyperfine coupling constants, and the chromium dimer. The size-consistency and size-extensivity of the methods are also discussed.
\end{abstract} 

\maketitle

\section{Introduction}
An elusive goal in electronic structure theory is to achieve a balanced description of dynamical and static correlation effects in a cost-effective manner. It is widely accepted that, while coupled-cluster theory\cite{Cizek66,Bartlett07,Shavitt09}  (CC) has been remarkably successful in capturing accurate dynamical correlation, the use of traditional single-reference (SR) methods is an ill-posed route for static correlation.\cite{Bulik15,Degroote16} This is because a reference wave function, e.g., Hartree--Fock (HF), is inadequate for strongly correlated systems, and the exponential wave operator $e^{\hat T}$ in SR-CC has to be improved by including higher excitation ranks, entailing a rapid increase in computational cost. This difficulty can be ameliorated by adopting some multiconfiguration (MC) {\it ansatz}, for which a complete active space self-consistent field\cite{Roos80,Siegbahn80} (CASSCF) is widely employed;  however, the formulation of multireference (MR)-CC is far less straightforward.\cite{Lyakh11,Kohn12}

From this point of view, symmetry-projected HF (PHF) presents an interesting paradigm,\cite{Scuseria11,Jimenez12}  because its wave function $\hat {\cal P} |\Phi\ket$ holds an MR structure, but can simultaneously be viewed as an SR method by regarding the projection operator $\hat {\cal P}$ as some kind of propagator acting on a Slater determinant $|\Phi\ket$. Motivated by this observation, we have recently explored alternative paths to handle static correlation using spin-symmetry projected unrestricted HF (SUHF) as a reference,\cite{Tsuchimochi14,Tsuchimochi15A,Tsuchimochi15C,Tsuchimochi16A,Tsuchimochi16B}  a subset of PHF that is equivalent to L\"owdin's spin-extended HF.\cite{Lowdin55B,Mayer73, Mayer80} The dynamical part of the correlation energy is then treated by either perturbation theory\cite{Tsuchimochi14} or configuration interaction (CI).\cite{Tsuchimochi16A,Tsuchimochi16B} Among these developments, it was shown that spin-extended CI with single and double substitutions (ECISD) can produce accurate results for systems where the static correlation plays an important role.
The success of this method is a result of its effective MR-CI nature, which comes from the MC character of SUHF. However, size-consistency and size-extensivity, which are of crucial importance in electronic structure theory,\cite{Bartlett89}  are both missing from ECISD (and MR-CI) as a consequence of the truncated CI.
In both SR-CI and MR-CI, the introduction of size-consistent corrections by Davidson \cite{Davidson74,Langhoff74} and others\cite{Pople77, Meissner88} partially mitigates the size-inconsistency error by approximating quadruple excitations.\cite{Duch94}  The generalization of the Davidson correction to ECISD (ECISD+Q) was shown to drastically improve the accuracy of ECISD; nonetheless, ECISD+Q is neither size-extensive nor variational with respect to the CI coefficients. The lack of the latter property is also troublesome, because the analytical gradient of ECISD+Q is not straightforward to compute, hindering the geometric optimization and computation of molecular properties (i.e., density matrix).\cite{Tsuchimochi17A}
Therefore, it is strongly desired that ECISD be extended such that (1) size-consistency and size-extensivity are established, and (2) analytical gradients remain readily available.
This is the purpose of the present work.

To accomplish size-extensive results, one can resort to a diagrammatic analysis to remove unlinked diagrams, as is routinely practiced in SR methods.\cite{Shavitt09}
However, the nonorthogonality involved in the spin-projection of SUHF poses a challenge, because a vacuum is not well defined (a problem shared by the MR formalism).
Simply removing the unlinked terms in the resulting ECISD equations might guarantee the size-extensivity of the method, but in the presence of a projection operator, this task seems particularly difficult. 

Besides the diagrammatic approach as well as the Davidson correction, in conventional electronic structure theory, several authors have modified the regular CI equations in an attempt to approximately cancel out the unlinked terms, retaining only the singles and doubles excitation manifold. The earliest methods in this spirit include coupled electron pair approximation (CEPA), which had been extensively developed in the SR paradigm until the advent of CC theory.\cite{Meyer71,Meyer73,Ahlrichs79,Ahlrichs85,Ahlrichs87,Daudey93} Although the accuracy of the SR-CEPA methods is by no means inferior to CCSD,\cite{Ahlrichs85,Ahlrichs87,Wennmohs08} they have lost popularity. CCSD is a more ``complete'' theory, as it is hierarchically improvable to the full CI (FCI) limit\cite{Kallay00,Hirata00} and is invariant with respect to the unitary rotation among the occupied space.
However, CEPA methods have attracted considerable attention for MR cases, because they are much simpler than MR-CC and a clearly better alternative to MR-CI, generically yielding a reliable description of dynamical correlation. 

With these past developments in mind, in the present work, we extend ECISD by adopting CEPA-like corrections for unlinked terms within the manifold of spin-projection. The proposed spin-extended CEPA (ECEPA) is an approximately size-consistent generalization of ECISD, and is thus expected to be more accurate. It is also considered to be an approximation whose accuracy lies between those of SR- and MR-CEPA, as the reference SUHF itself falls into the category of MC-SCF with far lower computational demands than conventional approaches. The trade-off is that the obtained energy is no longer the upper bound of the FCI energy. In this article, we will show how such a generalization may be realized within the overcomplete space of projected CI, and will report illustrative calculations to benchmark its performance. In addition, we will investigate the size-consistent and size-extensive properties using non-interacting Be atoms, and demonstrate the improved performance of ECEPA over ECISD and ECISD+Q.

Note that, although ECEPA is size-extensive in the sense that the correlation energy correctly scales with the number of electrons, the presence of a projection operator will not, in principle, permit exact size-consistency. While this could certainly limit the application range of ECEPA, we must stress that spin-projection is routinely applied in Kohn--Sham density functional theory\cite{Hohenberg64,Kohn65} (DFT) through Noodleman's approximation\cite{Noodleman81,Noodleman86} or Yamaguchi's weighted-average approximation,\cite{Yamaguchi86,Yamaguchi88} both of which remove spin-contamination only at first order. These methods yield encouraging results in various cases,\cite{Yamanaka94, Nishino97, Saito10} although it is easy to show that they are not size-consistent in general. From this perspective, while aiming for the same goal as these approximations, our methodology provides a more rigorous approach that is correct to the infinite order of $\hat S^2$. To this end, we will compare our method with Yamaguchi's scheme in terms of the accuracy of singlet-triplet splitting gaps.
It is shown that ECEPA gives encouraging results when approximately spin-projected DFT fails.

The organization of this paper is as follows. We describe the theoretical foundation of ECEPA in Section \ref{sec:ecepa}. We then introduce the concept of pair-energy and propose several size-consistent variants of ECISD in Section \ref{sec:pair-energy}, analogously to traditional CEPA approaches. Section \ref{sec:open-shell} deals with open-shell cases, which require some treatment such that ECEPA can be successfully reduced to SR-CEPA in the no spin-projection limit. 
We investigate the numerical performance of ECEPA on several test cases in Section \ref{sec:Results}; investigations include the potential energy curves of bond-breaking molecules (\ref{sec:dissociation}), singlet-triplet splitting gaps (\ref{sec:ST}), hyperfine coupling constants (\ref{sec:HFCC}), and Cr$_2$ (\ref{sec:Cr2}). The connection between ECISD+Q and ECEPA, along with their size-consistent properties, is studied in Section \ref{sec:Discussions}. Finally, we present our conclusions and outlook for future work in Section \ref{sec:Conclusions}.

\section{Theory}

\subsection{ECEPA equation}\label{sec:ecepa}
We consider the projected FCI wave function as
\begin{align}
\hat {\cal P}|\Psi\ket = \hat {\cal P}(1+\hat C)|\Phi_0\ket, \label{eq:PPsi}
\end{align} 
where $|\Phi_0\ket$ is the underlying determinant of SUHF.
By representing combinations of occupied and virtual orbital indices as $\mu,\nu,\tau$, the excitation operator $\hat C$ can be written as
\begin{align}
\hat C = \sum_\nu c_\nu \hat E_\nu,
\end{align}
where $c_\nu$ and $\hat E_\nu$ are the CI coefficients and excitation operators, respectively.
Hereinafter, we use $i,j,k,l$ to denote occupied, $a,b,c,d$ to denote virtual, and $p,q,r,s$ to denote general spin-orbitals (e.g., $|\Phi_\nu\ket = \hat E_\nu |\Phi_0\ket = |\Phi_{ij}^{ab}\ket$).
The spin-projected determinants  ($\hat {\cal P}|\Phi_Q\ket$) are not mutually orthogonal, and therefore the projected CI space in Eq.~(\ref{eq:PPsi}) is overcomplete.
As such, $\hat {\cal P}|\Psi\ket$  is not intermediate-normalized (we only assume the SUHF wave function is normalized). 

The CI coefficients are usually determined by projecting the Schr\"odinger equation with $\bra \Phi_\mu|$, or equivalently with $\bra \Phi_\mu|\hat {\cal P}$, noting that $\hat {\cal P} = \hat {\cal P}^\dag = \hat {\cal P}^2$ and $[\hat {\bar H}, \hat {\cal P}] = 0$. However, in the following, we reformulate the ECISD equation such that its correlating subspace, spanned by the projected single and double configurations, is modified to restore the size-consistency and size-extensivity, while the reference space is unchanged. Hence, we adopt the following projections:
\begin{align}
&\bra\Phi_0|(\hat {\bar H} - E_c) \hat {\cal P} |\Psi\ket = 0,\label{eq:Ec1}\\
&\bra \tilde \Phi_\mu | (\hat {\bar H} -E_c) \hat {\cal P}| \Psi\ket  =0\;\;\;\;\; \forall \; \mu \ne 0, \label{eq:CIEquation}
\end{align}
where $E_c$ is the correlation energy and $|\tilde \Phi_\mu\ket \equiv \hat Q |\Phi_\mu\ket$ removes the reference state with 
\begin{align}
\hat Q = 1 - \hat {\cal P} |\Phi_0\ket \bra \Phi_0 | \hat {\cal P}. \label{eq:Q}
\end{align}
Although $\hat Q = \hat Q^\dag = \hat Q^2$, the commutativity with $\hat {\bar H}$ is not satisfied. Therefore, Eqs.~(\ref{eq:Ec1}) and (\ref{eq:CIEquation}) constitute a non-Hermitian eigenvalue problem, but we emphasize that the resulting coefficients (and thus energy) remain exact. 

Using Eq.~(\ref{eq:Ec1}), the correlation energy is given by
\begin{align}
E_c &= \frac{\bra \Phi_0 | \hat {\bar H} \hat {\cal P}|\Psi\ket }{\bra \Phi_0|\hat {\cal P} |\Psi\ket} = \frac{\sum_{\nu} c_\nu \bra \Phi_0 | \hat {\bar H} \hat {\cal P} | \Phi_\nu \ket}{1+ \sum_\nu c_\nu \bra \Phi_0|\hat {\cal P}|\Phi_\nu\ket},\label{eq:projectiveEc}
\end{align} 
where $\hat {\bar H} \equiv \hat H - E_{\rm SUHF}$ with $E_{\rm SUHF} = \bra \Phi_0|\hat H \hat {\cal P} |\Phi_0\ket$ and $\bra \Phi_0|\hat {\cal P}|\Phi_0\ket =1$. As will soon become clear, it is convenient to express $E_c$ in the following self-dependent form:
\begin{align}
E_c = \sum_\nu \bra \Phi_0 | (\hat {\bar H} - E_c ) \hat {\cal P}|\Phi_\nu\ket c_\nu. \label{eq:Ec}
\end{align}  

ECISD only takes single and double excitations in $\hat C$, and neglects higher excitation effects, which are responsible for size-consistency and size-extensivity. To include such effects approximately, we consider the projected CC wave function instead of FCI:
\begin{align}
\hat {\cal P}|\Psi_{\rm CC}\ket = \hat {\cal P}\exp(\hat C)|\Phi_0\ket, \label{eq:PCC}
\end{align}
where $\hat C$ is truncated at doubles for the present, but $\exp (\hat C)$ generates higher excitations. 

In the presence of a spin-projection operator, $\bra \Phi_0|$ can generally interact with any $\hat {\cal P}|\Phi_\nu\ket$ including triples and higher excitations. Therefore, the projected CI equations (Eq.~(\ref{eq:CIEquation})) and the correlation energy (Eq.~(\ref{eq:Ec})) are non-terminating. However, such interactions are expected to be sufficiently weak, and hence we will neglect the Hamiltonian and overlap couplings between $\hat {\cal P}|\Phi_\mu\ket$ and $\hat {\cal P}|\Phi_\nu\ket$ when $|\Phi_\mu\ket$ and $|\Phi_\nu\ket$ differ by more than two electrons. The negative consequence associated with this approximation is assumed to be negligible compared to the expected improvement over ECISD gained by introducing the linked disconnected triples and quadruples, as well as that over spin-unrestricted SR methods by projecting them onto the correct symmetry space. 

Although there are many disconnected contributions in CCSD because of several powers of $\hat C$, we follow the conventional CEPA derivation, which only considers the quadratic contributions.\cite{Wennmohs08}
For this term, we write
\begin{align}
&\frac{1}{2} \bra \tilde \Phi_\mu | (\hat {\bar H} - E_c)  \hat {\cal P} \hat C^2 | \Phi_0\ket \bre
&= \frac{1}{2} \sum_{\lambda\tau} \bra  \tilde\Phi_\mu | \hat {\cal P}(\hat {\bar H} - E_c) \hat {\cal P} \hat E_\tau \hat E_\lambda |\Phi_0\ket c_\tau c_\lambda \bre
&\approx \sum_{\nu\lambda} \bra  \tilde\Phi_\mu | \hat {\cal P}|\Phi_\nu\ket \bra \Phi_\nu |(\hat {\bar H} - E_c)\hat {\cal P} \hat E_\lambda|\Phi_\nu\ket c_\lambda c_\nu \label{eq:ST1} \\
&= \sum_{\nu}  \bra  \tilde\Phi_\mu | \hat {\cal P}|\Phi_\nu\ket c_\nu (E_c - R_\nu). \label{eq:Ten1}
\end{align}
where we have introduced the resolution of the identity and truncated at doubles in Eq.~(\ref{eq:ST1}), and also removed the terms where neither $\lambda$ nor $\tau$ fully coincides with $\nu$. Using the energy expression in Eq.~(\ref{eq:Ec}), we have defined the residual contribution $R_{\nu}$  as
\begin{align}
R_\nu &= \sum_\lambda \Bigl(\bra \Phi_0 |(\hat {\bar H} - E_c) \hat {\cal P} \hat E_\lambda |\Phi_0\ket \bre
&\qquad - \bra \Phi_\nu|(\hat {\bar H} - E_c)\hat {\cal P} \hat E_\lambda | \Phi_\nu\ket\Bigr) c_\lambda,
\end{align} 
which is dominated by the so-called exclusion-principle-violating (EPV) term
\begin{align}
R_\nu \approx  \sum_{\lambda\cup\nu} \bra \Phi_0 | (\hat {\bar H} - E_c ) \hat {\cal P} \hat E_\lambda |\Phi_0\ket, \label{eq:Rnu}
\end{align} 
as suggested by its SR limit.
It should be emphasized that it is possible to avoid the introduction of the EPV terms by evaluating the matrix elements of the above quadratic contribution rigorously. However, not only will this treatment entail the derivation of higher-order coupling terms, but the resulting equation will also become nonlinear. This possibility is beyond our current scope and will be addressed in a forthcoming publication. In the present work, therefore, we explore the applicability of Eqs.~(\ref{eq:Ten1}) and (\ref{eq:Rnu}) as a first step, thus retaining the simplicity of the working equations and as much similarity to both ECISD and conventional CEPA as possible.

Defining $\hat R$ as some non-Hermitian operator such that
\begin{align}
&\hat R \;\hat {\cal P} |\Phi_0\ket = 0,\\
&\hat R \;\hat{ \cal P} |\Phi_\nu\ket  = R_\nu \hat {\cal P} |\Phi_\nu\ket,
\end{align} 
we find, for $\mu \ne 0$,
\begin{align}
&\bra \tilde \Phi_\mu | (\hat {\bar H} -E_c) \hat {\cal P} |\Psi_{\rm CC} \ket \approx \bra \tilde \Phi_\mu | (\hat {\bar H} - \hat R) \hat {\cal P} |\tilde\Psi\ket = 0,\label{eq:ECEPA}
\end{align} 
where $|\tilde \Psi\ket$ is an effective wave function in the form of CISD expansion.
Eq.~(\ref{eq:ECEPA}) is referred to as the ECEPA equation. The necessity of using $\bra \tilde \Phi_\mu|$ instead of $\bra \Phi_\mu|$ should now be even clearer---without the $\hat Q$ projection, Eq.~(\ref{eq:ECEPA}) would contradict Eq.~(\ref{eq:Ec1}) because of the contribution of $\bra \Phi_0|\hat {\cal P}$ in $\bra \Phi_\mu|$. 

The ECEPA equation  is therefore formally similar to the ECISD equation (Eq.~(\ref{eq:CIEquation}) with $|\Psi\ket \rightarrow |\tilde \Psi\ket$), and is easily cast as a generalized eigenvalue problem. Importantly, however, the former is independent of $E_c$, unlike the latter, which is manifested in the vast majority of unlinked diagrams being canceled out to reduce the size-inconsistency error in the former. Finally, Eq.~(\ref{eq:ECEPA}) rigorously reduces to the SR-CEPA problem if $\hat {\cal P} = \hat 1$, which we claim is an important property.

\subsection{Pair-energy and several approximations}\label{sec:pair-energy}
In this section, we derive several variants for $R_\nu$ as formally defined in Eq.~(\ref{eq:Rnu}). In the following, for simplicity, we discuss the case where $\hat C$ is composed of only double excitations. As in the standard CEPA variants, it is convenient to introduce pair energies $\epsilon_{ij}$, which divide the correlation energy into the contributions of each electron pair\cite{Szabo}:
\begin{align}
E_c = \sum_{i>j} \epsilon_{ij}.
\end{align} 
Given Eq.~(\ref{eq:Ec}), our definition of the pair-energy is
\begin{align}
\epsilon_{ij} = \sum_{a>b} \bra \Phi_0 |\left(  \hat {\bar H}  - E_c\right) \hat {\cal P} | \Phi_{ij}^{ab} \ket c_{ij}^{ab}.
\end{align}
Neglecting the dependence of virtual indices in $R_{\nu}$, as in most regular CEPA methods, the EPV terms are approximated as
\begin{align}
R_{ij}^{ab} \approx \sum_{k>l}^{(ij)} \epsilon_{kl}. \label{eq:EPV}
\end{align}
 
We are now ready to set out several variants of CEPA with different treatments of Eq.~(\ref{eq:EPV}), by analogy of the single reference methods. If all the EPV terms $R_{ij}^{ab}$ are completely neglected, we obtain a large cancellation between the unlinked terms in Eq.~(\ref{eq:ECEPA}). This is historically called CEPA(0),\cite{Ahlrichs79,Koch81}  and is equivalent to linearized CC theory (LCC).\cite{Bartlett1981,Laidig84,Laidig87,Taube09} Hence, we will refer to this approximation as ELCC.
Despite a crude approximation, LCC is known to be more accurate than full CC for some cases.\cite{Taube09}
It is worth noting, however, that ELCC is not exact for two-electron systems, adding some nonzero correction to ECISD although the latter is already exact.

One should deal with the EPV terms appropriately to recover the exactness for two-electron systems. The resemblance between pair-energies in SR and spin-projected approaches permits us to introduce ECEPA(1-3)\cite{Kelly63,Kelly64,Meyer73,Meyer74,Koch81} :
\begin{subequations}\begin{align}
&R_{ij}^{ab} = \epsilon_{ij} + \frac{1}{2} \sum_{k\ne i,j} (\epsilon_{ik} + \epsilon_{kj}), & \rm  [ECEPA(1)]\\
&R_{ij}^{ab} = \epsilon_{ij}, & \rm [ECEPA(2)]\\
&R_{ij}^{ab} = \epsilon_{ij} + \sum_{k\ne i,j}  (\epsilon_{ik} + \epsilon_{kj}), & \rm [ECEPA(3)]
\end{align} \end{subequations}
Note that these approximations are not invariant with respect to an orbital rotation among occupied orbitals. As such an orbital dependence precludes a straightforward derivation of the energy functionals and analytical gradients, we will not consider these approaches further in the present study.

To avoid the orbital dependence in ECEPA(1-3), we follow Gdanitz and Ahlrichs in averaging the pair energies.\cite{Gdanitz87}
By partitioning the system into $n_e/2$ non-interacting electron pairs ($n_e$ is the number of correlated electrons), the average of the pair energies becomes
\begin{align}
\bar \epsilon 
&= \frac{2}{n_e} E_c. \label{eq:ACPF}
\end{align}
Then, the so-called averaged coupled-pair functional (ACPF) assumes that all EPV pairs are approximated by Eq.~(\ref{eq:ACPF}).\cite{Gdanitz87} Meissner suggested a different averaging by considering the actual number of pairs, $\tiny\begin{pmatrix} n_e \\ 2\end{pmatrix}$.\cite{Meissner88}  The averaged pair-energy in this case becomes
\begin{align}
\bar \epsilon &
= \frac{4n_e-6}{n_e (n_e -1 )} E_c. \label{eq:AQCC}
\end{align}
Szalay and Bartlett employed Eq.~(\ref{eq:AQCC}) for MR calculations, and derived the averaged quadratic coupled-cluster (AQCC).\cite{Szalay93,Fusti96}
In accordance with these previous developments, we applied these averaging schemes to our ECISD generalization, giving EACPF and EAQCC, respectively. 

The accuracy of these methods can only be assessed by actual calculations.  However, as will be argued in Section  \ref{sec:Results}, it turns out that EACPF and EAQCC have different advantages and disadvantages. Namely, the ACPF parametrization is more accurate when it is stable, but can overestimate the higher excitation effects, whereas that of AQCC is generally more stable at the cost of slight underestimation of dynamical correlation effects. Interestingly, exactly the same observation is made in conventional MR adaptations.\cite{Szalay93,Szalay95,Gdanitz01}
Hence, it is indicative that neither Eq.~(\ref{eq:ACPF}) nor (\ref{eq:AQCC}) is conclusive for the average treatment of EPV terms. Considering that these averaging schemes are themselves approximations to the orbital-dependent EPV treatments ECEPA(1) and ECEPA(3),\cite{Gdanitz01,Wennmohs08} we also propose a compromised scheme by taking a linear combination as
\begin{align}
R_{ij}^{ab} = \left((1-a) \frac{2}{n_e} + a \frac{4n_e-6}{n_e (n_e -1 )} \right)E_c,\label{eq:AQCCh}
\end{align} 
with an empirical parameter $a$. In the following, we set $a= 0.65$, which was found to provide suitable compensation for the undesired errors in EACPF and EAQCC. We will refer to this scheme as EAQCC hybrid (EAQCCh). Note that EACPF, EAQCC, and EAQCCh are all exact for two-electron systems, reducing to ECISD.
 
Finally, we note that there can be redundancy terms in these methods, as in MR-CEPA,\cite{Ruttink91,Malrieu94}  arising from the overcompleteness of the projected CI space. However, such redundancy is completely removed by taking the metric into account.

\subsection{Open-shell cases}\label{sec:open-shell}
For open-shell systems that include unpaired electrons, we also follow the treatment of the original ACPF and AQCC approaches, so that EACPF and EAQCC coincide with their SR limits whenever $\hat {\cal P} = \hat 1$. In this subsection, we will briefly discuss how we achieve this goal.

In both the SR and MR variants, the orbital space is divided into three subspaces: doubly-occupied inactive, active (open-shell for SR), and external orbitals. Among all possible configurations considered in CISD, cluster corrections are not needed for the internal configuration space, where external orbitals remain unoccupied. For the ECEPA methods, similarly, we need to define an appropriate internal space to be removed from the EPV treatment. Here, as in the SR case, we define our internal space to include the reference SUHF and the projected configurations that substitute electrons into open-shell orbitals, which can be uniquely identified by diagonalizing the reference density matrix.\cite{Tsuchimochi10B,Tsuchimochi11}  Removing the internal space in Eq.~(\ref{eq:ECEPA}) guarantees that EACPF and EAQCC can be safely reduced to their original formulations for an ROHF reference when no projection operation is carried out. Evidently, this can be accomplished by generalizing $\hat Q$ in Eq.~(\ref{eq:Q}), but would formally require the inverse of the overlap matrix $S_{\mu\nu} = \bra \Phi_\mu| \hat {\cal P}|\Phi_\nu\ket$ within the internal space ($\mu,\nu \in {\rm int}$). Nevertheless, it is worth mentioning that $S_{\mu\nu}$ is easily shown to be {\it diagonal} in the corresponding-pair orbital (CO) basis. Therefore, instead of Eq.~(\ref{eq:Q}), we use
\begin{align}
\hat {Q} = 1 - \sum_{\mu \in \rm int} \frac{\hat {\cal P}|\Phi_\mu\ket \bra \Phi_\mu| \hat {\cal P}}{\bra \Phi_\mu | \hat {\cal P}|\Phi_\mu\ket}.\label{eq:generalQ}
\end{align}
Note that Eqs.~(\ref{eq:Q}) and (\ref{eq:generalQ}) are identical to one another for closed-shell systems.

\subsection{Dressed Hamiltonian and energy functional}\label{sec:EnergyFunctional}
As mentioned, the above ECEPA approximations can be recast as an eigenvalue problem by shifting the Hamiltonian matrix in the ECISD equation.\cite{Heully92} The matrix representation of Eq.~(\ref{eq:ECEPA}) is shown to be
\begin{align}
{\bf H} {\bf c} = E_c {\bf S} {\bf c},\label{eq:HcESc}
\end{align} 
where
\begin{align}
& {H}_{\mu\nu} = \bra \Phi_\mu |\hat {\bar H} \hat  {\cal P}| \Phi_\nu\ket + \bra \Phi_\mu| \hat {\cal P} \hat {Q}|\Phi_\nu\ket (E_c- R_\nu),\\
& S_{\mu\nu} = \bra \Phi_\mu |\hat  {\cal P}| \Phi_\nu\ket.
\end{align} 
The second term is the size-consistent correction that shifts the Hamiltonian by $E_c - R_\nu$ under the metric of the projected manifold.

For the averaged models introduced in Section \ref{sec:pair-energy}, including ELCC, the unitary invariance with respect to orbital rotation permits us to define an effective (total) energy functional as 
\begin{align}
 F[{\bf c}] = \frac{\bra \tilde \Psi | \hat H \hat {\cal P} |\tilde \Psi\ket - \zeta E_{\rm SUHF}\bra \tilde \Psi | \hat{\cal P}\hat{Q} |\tilde\Psi\ket }{\bra \tilde \Psi |\hat {\cal P} |\tilde \Psi\ket - \zeta \bra \tilde \Psi|\hat{\cal P}\hat{Q}|\tilde \Psi\ket },\label{eq:Fc}
\end{align} 
where $\zeta$ is a constant depending on the method; $\zeta = 0, 1, (n_e-2)/n_e$, and $(n_e-2)(n_e-3)/n_e(n_e-1)$ for ECISD, ELCC, EACPF, and EAQCC, respectively. {\bf c} is variationally determined by minimizing $F[{\bf c}]$; one can easily show that requiring $\dd F[{\bf c}]/\dd {\bf c} = {\bf 0}$ results in Eq.~(\ref{eq:HcESc}).

Size-extensivity is guaranteed because the residual correlation $E_c$ is known to be size-extensive in the SR and MR approaches, even though the correlation energy of SUHF ($\Delta E_{\rm SUHF} = E_{\rm SUHF} - E_{\rm HF}$) is not.  However, size-consistency is not rigorously satisfied in general, but only for special cases, because of the presence of $\hat {\cal P}$. We will discuss these aspects in Section \ref{sec:SC}.

\subsection{Density matrix and nuclear gradient}
The variationality of Eq.~(\ref{eq:Fc}) guarantees the facile evaluation of the analytical derivative of the proposed methods by generalizing the procedure used in the ECISD energy derivative.\cite{Tsuchimochi17A,Helgaker89,Jorgensen88}  Here, we briefly note how one can derive the relaxed density matrix and nuclear gradient of averaged ECEPA.

The {\it unrelaxed} density matrix can be obtained by taking the linear response of $F$ with respect to an infinitesimal one-body perturbation in the Hamiltonian, without taking into account the orbital relaxation effect.\cite{Ahlrichs85,Szalay00}   This results in a mixture of ECISD and SUHF density matrices:
\begin{align}
{\bf P} =   \frac{ {\bf P}^{\rm ECISD} -\zeta\bra \tilde\Psi|\hat{\cal P}\hat{Q}|\tilde\Psi\ket {\bf P}^{\rm SUHF} }{1 -\zeta\bra \tilde\Psi|\hat{\cal P}\hat{Q}|\tilde\Psi\ket}. \label{eq:1PDM}
\end{align}
where we have assumed $\bra \tilde \Psi| \hat {\cal P} |\tilde \Psi\ket = 1 $. 
The expression for the relaxation correction term ${\bf P}^{\rm corr}$ is unchanged from that of ECISD,\cite{Tsuchimochi17A}  as such a term is understood as the orbital gradient of the density matrix of the reference wave function, i.e., SUHF. By parametrizing the molecular orbitals {\bf C} as 
\begin{align}
{\bf C} = {\bf C}_0 e^{\bm \kappa},
\end{align}  
with ${\bm \kappa}$ being an anti-Hermitian matrix, we have previously shown that
\begin{align}
{\bf P}^{\rm corr} =\sum_{ai} z_{ai} \Bigl(  \frac{\dd}{\dd \kappa_{ai}^*} +  \frac{\dd }{\dd \kappa_{ia}^*}\Bigr) {\bf P}^{\rm SUHF} \Biggr|_{{\bm \kappa} = {\bf 0}}.\label{eq:Prel2} 
\end{align}
Here, {\bf z} is the solution of the coupled-perturbed SUHF equation, which requires the orbital derivative of $F$ ($\tilde L_{pq}$). Given Eq.~(\ref{eq:Fc}), it is straightforward to show that
\begin{align}
\tilde L_{pq} = \frac{\dd F}{\dd \kappa_{pq}^*} =  \frac{L_{pq} + \zeta E_c \; \xi_{pq}} {1 - \zeta \bra \tilde\Psi|\hat{\cal P}\hat{Q}|\tilde \Psi\ket},
\end{align} 
where $L_{pq}$ and  $\xi_{pq}$ are the orbital gradients of $E_{\rm ECISD}$ (i.e., $F[\zeta=0]$) and $\bra \tilde \Psi | \hat{\cal P}\hat{Q} |\tilde \Psi\ket$. The derivation of the former is given in Ref.[\onlinecite{Tsuchimochi17A}] with more detailed discussions. The latter can be conveniently expressed as the density matrix of a pseudo-wave function $|\Psi'\ket$:
\begin{align}
\xi_{pq} = \frac{\dd \bra \tilde\Psi|\hat{\cal P}\hat{Q} | \tilde\Psi\ket}{\dd \kappa^*_{pq}} 
&=\bra \Psi' | \hat E_{qp}\hat {\cal P}|\Psi' \ket,
\end{align} 
where $|\Psi'\ket$ has the same structure as $|\tilde\Psi\ket$, but with the following modified CI coefficients:
\begin{align}
c_\nu' =
\begin{cases}
c_\nu - S_{\nu\nu}^{-1} \sum_\mu S_{\nu\mu} c_\mu & (\nu \in { \rm int})\\
c_\nu & ({\rm otherwise})
\end{cases}.
\end{align} 
Obviously, the coupled-perturbed SUHF equation is now to be solved with $\tilde {\bf L}$ instead of ${\bf L}$, which then gives rise to a different {\bf z} from that of ECISD. This is where the chief difference between ECEPA and ECISD comes in. The relaxed density matrix ${\bf P}^{\rm rel} = {\bf P} + {\bf P}^{\rm corr}$ so obtained is not spin-adapted, because we started from Eq.~(\ref{eq:Fc}), which projects only the ket state. The spin-adaption of ${\bf P}^{\rm rel}$ is performed using the Wigner--Eckart theorem, which gives the correct relaxed spin-density matrices for $\alpha$ and $\beta$ spins.\cite{Tsuchimochi17A,JJSakurai}

Other critical differences from the ECISD gradient lie in the explicit derivative of $F$ with respect to nuclear coordinate $x$, as well as in the energy-weighted density matrix, both of which can nonetheless be given as some linear combination of the corresponding ECISD and SUHF quantities, as in Eq.~(\ref{eq:1PDM}). The interested reader is referred to the supplemental material for their complete expressions. 

\section{Computational details}
The proposed schemes have been implemented in the GELLAN suite of programs. We have used the integral form of the projection operator, which is computationally more convenient to deal with than L\"owdin's operator.\cite{RingSchuck,Scuseria11,Jimenez12}  For all the calculations reported in this article, the number of grid points $N_{\rm grid}$ for the numerical integrations was set to either 3 or 4, unless otherwise noted, which was found to be sufficient to obtain the desired $\bra \hat S^2 \ket$ value with numerical precision of $10^{-9}$. This means that our methods are effectively multireference; a reference SUHF state is represented by three or four nonorthogonal Slater determinants.

Most of the calculations reported in this paper do not correlate core electrons. To perform a frozen-core calculation in ECISD and ECEPA, we employed the constrained optimization in SUHF to obtain the HF core orbitals. \cite{Tsuchimochi16B}  This can be achieved by averaging the core-virtual (CV) block (and VC block) of the SUHF Fock matrix as
\begin{align}
{\bm\calF'}_{ \rm CV}^\sigma = \frac{1}{2}\left( {\bm\calF}^{\alpha}_{ \rm CV} + {\bm\calF}^{\beta}_{ \rm CV} \right),
\end{align} 
for both $\sigma = \alpha, \beta$. At SCF convergence, it is guaranteed that the $\alpha$ and $\beta$ orbitals share the same core space,\cite{Tsuchimochi10A,Tsuchimochi10B,Tsuchimochi11}
while the SUHF energy is minimized under such a constraint. 

\tabcolsep=2mm
\begin{table}
\caption{NPE ($\Delta_{\rm max} - \Delta_{\rm min}$) of dissociation curves for several molecules in m$E_h$.}\label{tb:NPE}
\begin{tabular}{lrrrrrrr}
\hline\hline
Method	& 	 HF 	& 	F$_2$	&	H$_2$O 	& 	N$_2$ 	& 	C$_2$	&	Mean\\
\hline
UCISD	&	19.0	&	32.3		&		35.6	&	59.4		&	39.8		&	37.2\\
UCCSD	&	6.1	&	11.8		&		13.2	&	24.3		&	23.0		&	15.7\\
UCCSD(T)&	3.4	&	6.7		&	8.6	&	15.4		&		27.1		&	12.2\\
SUHF 	&	14.0 &	20.7		& 		45.3	&	104.0	&	43.8		&	45.6\\
ECISD 	&	0.9 	&	2.4		&		3.8	&	15.6		&	9.9		&	6.5\\
ECISD+Q &	0.1	&	0.2	 	&   		1.3	&	1.9		&	3.6		&	1.4\\
ELCC 	&	0.2	&	1.5		&		3.2	&	16.9		&	20.5		&	8.5\\
EACPF	&	0.2	&	0.2		&		1.8	&	6.8		&	5.0		&	2.8\\
EAQCC	&	0.3	&	0.5		&		1.1	&	3.9		&	4.7		&	2.1\\
EAQCCh  &	0.3	&	0.4		&		1.1	&	3.4		&	3.8		&	1.8\\
\hline\hline
\end{tabular} 

\end{table}

\section{Illustrative calculations}\label{sec:Results}
\subsection{Molecular dissociation}\label{sec:dissociation}
We first investigate the accuracy of the proposed ECEPA variants by computing the bond-dissociation curves of several molecules that include HF, F$_2$, H$_2$O, N$_2$, and C$_2$. For H$_2$O, both OH bonds are symmetrically stretched with the angle fixed to 109.57$^\circ$. Here, we use a small 6-31G basis set and freeze $1s$ orbitals in the homonuclear diatomic molecules for comparison with FCI.  As an estimate of the accuracy of each method, Table \ref{tb:NPE} summarizes the non-parallelity errors (NPE) in m$E_h$, computed by taking the difference between the maximum and minimum deviations ($\Delta_{\rm max}$ and $\Delta_{\rm min}$) from the FCI energy across the reaction coordinate. 

\begin{figure}
\includegraphics[width=85mm, bb=0 0 279 337]{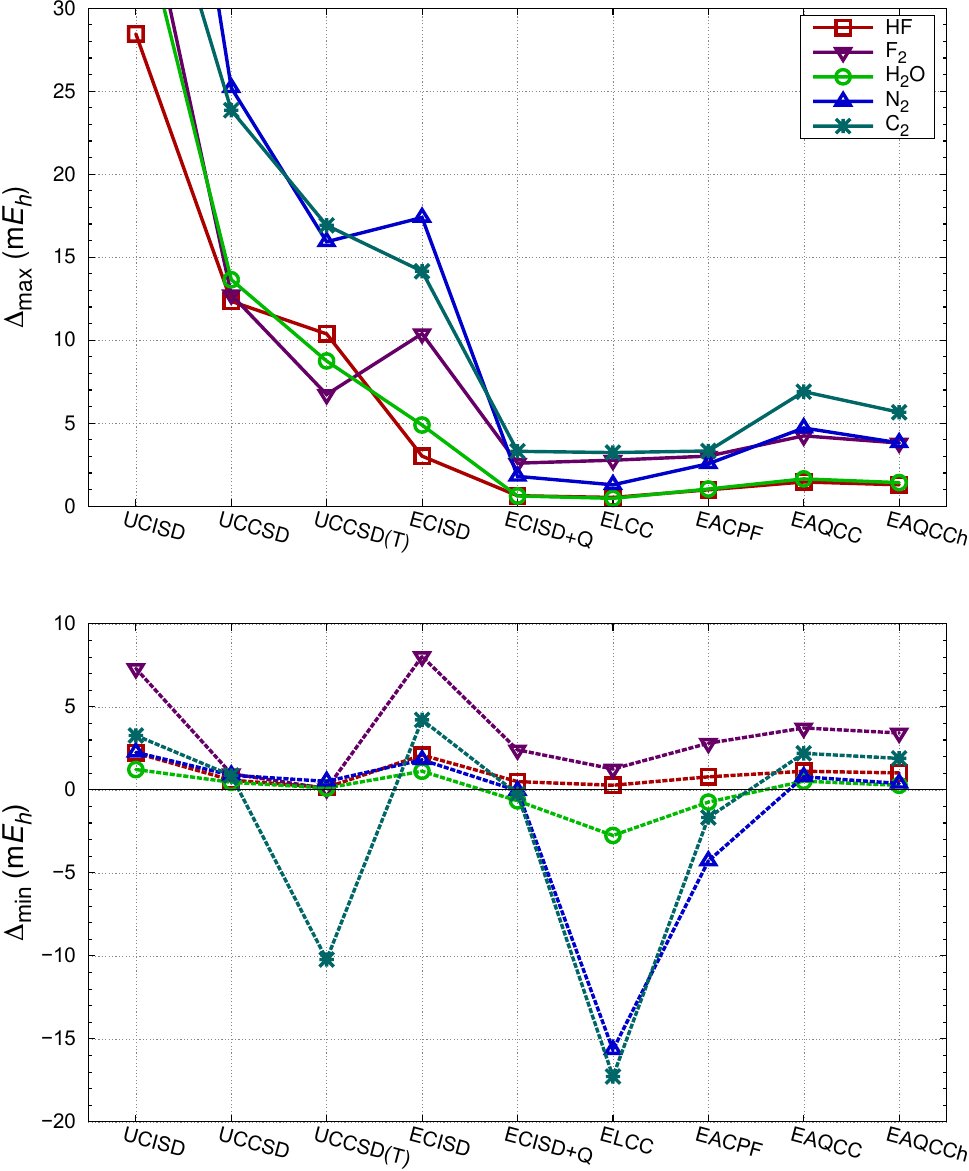}
\caption{Maximum (top) and minimum (bottom) differences from FCI energy for the tested dissociation profiles.}\label{fig:MaxMin}
\end{figure}

It is well known that spin-unrestricted (U) calculations serve as an easy means of obtaining bond-dissociation curves of molecules, and are indeed actively employed in density functional applications. However, it is also widely accepted that, as can be seen in the results of UCISD, UCCSD, and UCCSD(T), they incur large errors due to spin-contamination, which necessarily pushes up the ground state energy by mixing different spin states, i.e., excited states. In Figure \ref{fig:MaxMin}, we plot $\Delta_{\rm max}$ (top panel) and $\Delta_{\rm min}$ (bottom panel) for each dissociation curve.
Because unrestricted methods usually dissociate molecules to the energetically correct limits, where they are considered to be most accurate, $\Delta_{\rm min}$ are acceptably small for these methods. The only exception is C$_2$/UCCSD(T), which tends to diverge in the intermediate region of the dissociation path.
The large $\Delta _{\rm max}$ values in Figure \ref{fig:MaxMin} manifestly show that spin-contamination causes the correlation energy to be underestimated.

Whereas our spin-extended methods ECISD and ECISD+Q reduce these errors by removing spin-contamination, the results of ECEPA methods largely depend on how the EPV terms are treated.
ELCC completely ignores the EPV terms, and therefore severely overestimates the quadruple and higher excitation effects. 
This results in the substantial overestimation of the correlation energy, as can be clearly seen in the bottom panel of Figure \ref{fig:MaxMin}.
Moreover, we note that the $\Delta_{\rm max}$ values given by ELCC are very similar to those of ECISD+Q and EACPF, which are all sub-m$E_h$ errors away from FCI. Therefore, the large NPEs of ELCC are solely a consequence of the overemphasized effect of cluster excitations, especially in multiple bond-breaking of N$_2$ and C$_2$. 

EACPF accounts for the EPV effects in an average manner within the independent pair model.
As a result, it suppresses the overcorrelation effect by removing most of the undesired excitations in the cluster expansion, and yields much better results compared to ELCC; the average NPE is reduced from 8.5 to 2.8 m$E_h$.
However, it appears that EACPF gives an insufficient cancellation of cluster excitation effects in N$_2$ and C$_2$. In contrast, EAQCC treats the EPV terms more satisfactorily than EACPF (and ELCC), and yields very small $\Delta_{\rm min}$. In turn, the $\Delta_{\rm max}$ values of EAQCC become a little larger, meaning that higher excitation effects may be undershot in some cases. Consequently, EAQCC provides NPEs that are similar to, yet slightly better than, those of EACPF on average. These results strongly imply that it is crucial to handle the EPV terms appropriately for a method to be accurate. The hybrid model, EAQCCh, combines the benefits of the two approaches; it diminishes both $\Delta_{\rm max}$ and $\Delta_{\rm min}$ compared with EAQCC for all the tested molecules, resulting in a further improved description of dissociation curves with a mean NPE of 1.8 m$E_h$.

Though EAQCC and EAQCCh are successful overall, the {\it ad hoc} treatment of size-consistency correction in ECISD+Q appears to be surprisingly accurate (1.4 m$E_h$ error).
However, it should be stressed that the superior performance of ECISD+Q is in part attributed to an error cancellation, as will be analyzed in Section \ref{sec:Taylor}.

\begin{table}
\caption{Mean errors and mean absolute errors from experimental values$^{\rm a}$  of adiabatic singlet-triplet splitting gaps $T_e = E_{\rm S} - E_{\rm T}$ computed with aug-cc-pVQZ (kcal/mol). ``max.'' indicates the maximum deviation in $T_e$ and the corresponding molecule.}\label{tb:ST}
\tabcolsep=0.5mm
\begin{threeparttable}[t]
 \begin{tabular}{lrcrl}
\hline\hline
	&\qquad ME & {\qquad}  MAE  {\qquad}  & \multicolumn{2}{c}{max.} \\
\hline
SUHF & 	-5.8 &  6.2 &	-12.9& (O$_2$) \\
ECISD & -1.0 & 1.1 &	-5.4& (NF)\\
ECISD+Q & -0.8 & 1.3 & 	-4.6& (NF)\\
ELCC &	-0.8 & 1.7	&	-7.0& (CF$_2$)\\
EACPF &	-1.0 & 1.4	&	-6.0& (CF$_2$)\\
EAQCC & -1.0 & 1.2 & 	-5.3& (CF$_2$)\\
EAQCCh & -1.0 & 1.3 &  -5.5 & (CF$_2$) \\
CCSD\tnote{b} & 5.3 & 5.4 & 		11.8& (OH$^+$)\\
CCSD(T)\tnote{b}  & 3.0 & 3.2 &	7.3& (NH)\\
CASSCF\tnote{c,d} & 3.5 & 3.7 & 	9.2& (H$_2$CC)\\
CASPT2\tnote{c,d} & 1.5 & 2.7  &	8.6& (O$_2$)\\
PBE\tnote{c,e} & -3.4 & 7.8 & -14.7 & (OH$^+$)\\
tPBE\tnote{c,f} & -3.6 & 3.8 & 	-11.5& (OH$^+$)\\
W3X-L\tnote{c} & 0.4 & 1.0 &		2.6& (H$_2$CC)\\
\hline\hline
 \end{tabular}
 {\footnotesize
\begin{tablenotes}
\item[a] Ref.[\onlinecite{Bao16}] and references therein.
\item[b] Restricted and unrestricted HF orbitals for singlet and triplet states.
\item[c] Taken from Ref.[\onlinecite{Bao16}].
\item[d] Full-valence active space.
\item[e] Weighted average of broken-symmetry calculations based on the Yamaguchi formula.
\item[f] Active space chosen based on extended correlated participating orbitals.
\end{tablenotes}
}
 \end{threeparttable}
\end{table}

\subsection{Singlet-triplet splitting gaps}\label{sec:ST}
Spin-projection offers immediate access to different spin states by simply changing the designated quantum numbers in the projection operator.
This allows us to compute energy gaps between different spin states. Here, we investigate the performance of our methods on singlet-triplet splitting gaps $T_e = E_{\rm S}- E_{\rm T}$ using the test set that was recently used by Bao {\it et al}.\cite{Bao16} We will adopt the same computational conditions as in their work; the basis set is aug-cc-pVTZ and the geometries are optimized with QCISD(T). We have again used the frozen-core approximation for $1s$ orbitals. For the elements in the second row of the periodic table, $2s$ orbitals are additionally frozen. 

Before proceeding, we should point out that one disadvantage of using collinear spin-projection is that, depending on the multiplicity of the underlying broken-symmetry wave function, there are essentially $2s+1$ different {\it ans\"atze} that give the same total spin $s$. Ideally, all these wave functions should give the same energy for the non-relativistic Hamiltonian, but the approximate nature of SUHF as well as ECISD and ECEPA does not guarantee such equality (although some states automatically possess the same energy because of the time-reversal symmetry).
One way to overcome this problem is to extend these methods to the non-collinear regime, which unfortunately necessitates a substantial increase in computational cost.\cite{Jimenez12}
In this paper, we choose the low-spin states for triplet calculations rather than high-spin ones. We consider the use of low-spin triplet states to be more consistent in that they share the same multiplicity as the singlet states. One can expect some error cancellation by treating singlets and triplets similarly, rather than performing unpaired calculations for triplet states. For other methods, we adopt high-spin triplets.

In Table \ref{tb:ST}, we have collected the mean errors (ME) and mean absolute errors (MAE) of $T_e$ compared to experimental values, along with the largest deviation (max.). For CCSD and CCSD(T), RHF and UHF orbitals are used for singlets and triplets. In addition to our results, we also show the multireference results of CASSCF and its second-order perturbation theory\cite{Andersson90,Andersson92} (CASPT2) with full-valence active spaces. For the latter, an empirical level shift of 0.25 eV was adopted.\cite{Ghigo04} A comparison with DFT is made using the PBE functional,\cite{PBE} either with the weighted-average scheme of Yamaguchi\cite{Yamaguchi86,Yamaguchi88} or the multiconfiguration pair-density functional theory of Gagliardi and Truhlar (tPBE).\cite{Manni14,Carlson15,Bao16} The most accurate SR results are available with the W3X-L method, which is an approximation of CCSDT(Q) in the complete basis set (CBS) limit.\cite{Chan15,Bao16}

\begin{figure}[t!]
\includegraphics[width=85mm, bb=0 0 413 241]{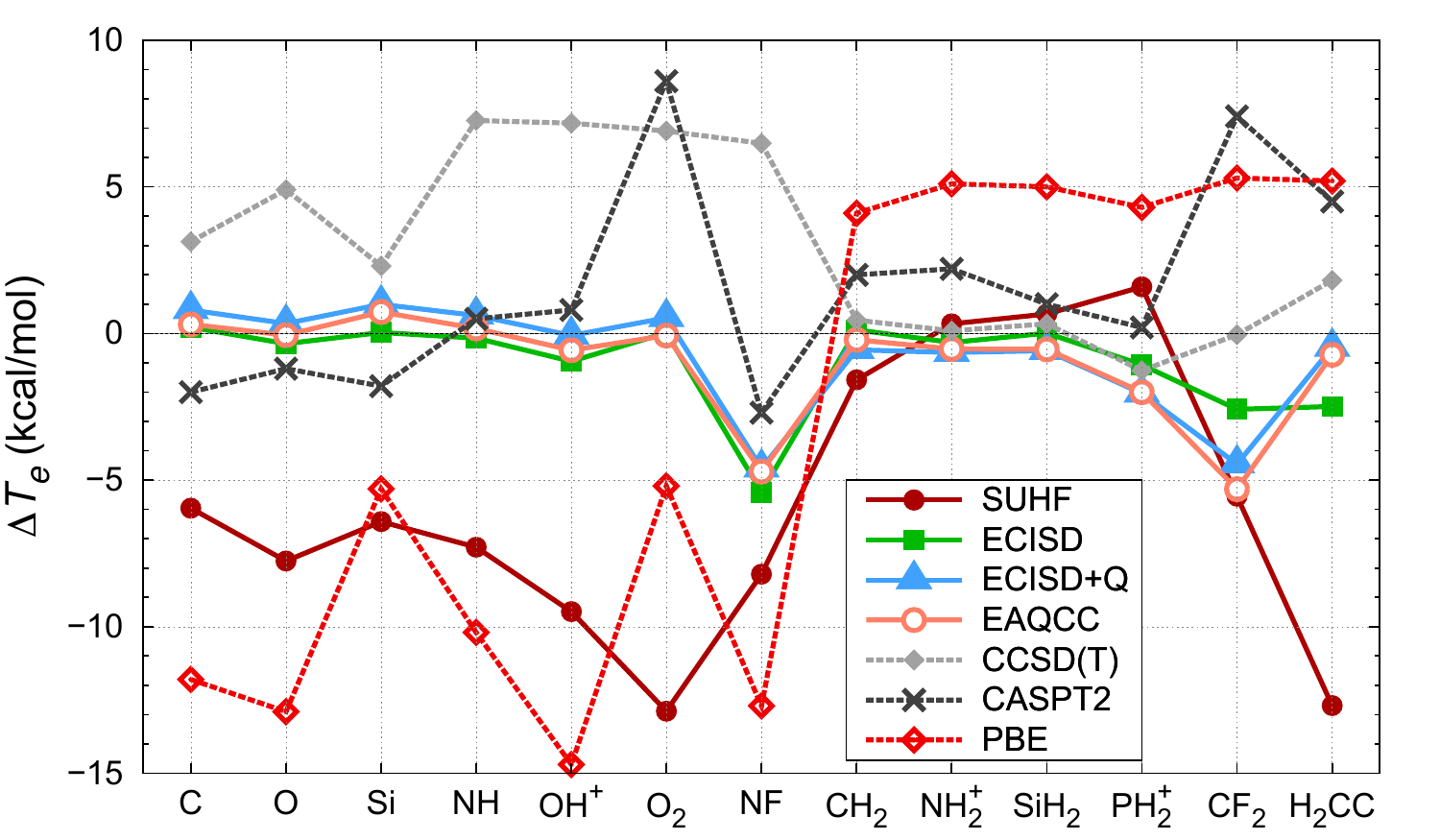}
\caption{Errors of $T_e = E_{\rm S} - E_{\rm T}$ from experimental values (kcal/mol) for chosen methods. }\label{fig:ST}
\end{figure}

As previously observed, SUHF is as accurate as CCSD for $T_e$, but for different reasons. The former incorporates static correlation but neglects dynamical correlation, making the correlation energy in singlet states more pronounced than in triplet states. In contrast, CCSD and CCSD(T) poorly describe static correlation effects, causing inaccurate results for biradical singlet molecules. This can be observed in Figure \ref{fig:ST}, where we have plotted the errors for the methods under consideration. In this figure, the molecules on the left side (C to NF) exhibit a prominent biradical character in their singlet states (strongly static), whereas wave functions are less multiconfigurational in those on the right side (CH$_2$ to H$_2$CC), although an appropriate treatment for static correlation is still crucial (weakly static). When singlet states are governed by strong static correlation, SUHF and CCSD(T) suffer from marked underestimation and overestimation of $T_e$ for the aforementioned reason. In contrast, when the singlets are weakly static, CCSD(T) behaves very well; this result implies restricted CCSD(T) is able to handle static correlation partially, although not perfectly by any means.
In passing, the use of unrestricted orbitals in CCSD(T) for singlets leads to a catastrophic failure in predicting $T_e$, especially when $E_{\rm T} < E_{\rm S}$; the resulting MAE is 13.7 kcal/mol (not shown in Table \ref{tb:ST}).

The performance of PBE with Yamaguchi's approximate spin-projection is also clearly affected by the strong and weak static correlation regimes.
Unfortunately, this approximation is not sufficiently accurate, with an MAE of 7.8 kcal/mol. It has recently been suggested that empirically ``tuning'' the weighted-average scheme with a scaling factor could systematically improve the DFT results for spin gaps.\cite{Ko13,Cho15} Indeed, with the proposed scaling factor for PBE (1.392),\cite{Cho15} the MAE of $T_e$ is improved to 5.3 kcal/mol. tPBE instead employs a CAS reference, being a clearly better alternative to these schemes with a reduced MAE of 3.8 kcal/mol. However, the active orbitals must be carefully selected to make the method accurate. Additionally, the MAE of CASSCF is 3.7 kcal/mol, meaning that the tPBE correction simply worsens $T_e$. The overestimation of $T_e$ in CASSCF is mainly attributed to the missing dynamical correlation that involves external orbitals. The perturbative correction in CASPT2 improves CASSCF, albeit only slightly, by 1 kcal/mol in MAE. Again, as in tPBE, the active space must be thoughtfully chosen to account for balanced pair-correlations. Note that if the active space was the same as for tPBE, the MAE of CASPT2 would be reduced to 0.8 kcal/mol.\cite{Bao16} 

The ECISD and ECEPA methods, without specifying active spaces, give highly accurate $T_e$; the black-box characterization of strong static correlation in spin-extended methods is very appealing. We do not present the results for ELCC, EACPF, and EAQCCh in Figure \ref{fig:ST}, as they behave very similarly to that of EAQCC, although ELCC is slightly less accurate than the others. It turns out that ECISD, ECISD+Q, and EAQCC are all in good agreement, with similar accuracy to W3X-L. This strongly suggests that most of the size-inconsistency errors successfully cancel out when calculating $T_e$, and ECISD already captures a balanced description between singlet and triplet states for these test molecules. When high-spin triplet references are used, the change in $E_{\rm T}$ is found to be less than 1 m$E_h$ in most cases. However, occasionally, the high-spin and low-spin triplet energies can be significantly different; for H$_2$CC, the former energy is higher than the latter by 15--19 m$E_h$. Consequently, the results obtained with high-spin references are slightly worse; the MAEs of ECISD, ECISD+Q, and EAQCC are 2.6, 2.7, and 2.5 kcal/mol, respectively, which are nevertheless better than those of most other SR and MR methods.

\begin{table}
\caption{Errors in the computed hyperfine coupling constants in MHz from experimental reference values. All calculations were performed with the EPR-III basis set. ``div'' stands for divergence of calculation.}\label{tb:HFCC}
\begin{threeparttable}[t]
\begin{tabular}{lrrrrrrrrrrrrr}
\hline\hline
& \multicolumn{2}{c}{BO} & \multicolumn{2}{c}{CO$^+$} & \multicolumn{2}{c}{CN}\\
& 	$^{11}$B	&	$^{17}$O	&	$^{13}$C	& 	$^{17}$O	& $^{13}$C 	&	$^{14}$N\\
\hline
UHF		&	118	&	41	&	356	&	72	&	709	&	-27\\
SUHF	&	60	&	133	&	87	&	74	&	495	&	-20\\
\\
(Unrelaxed)\\
ECISD 	&	69	&	43	&	119	&	18	&	412	&	-8\\
ELCC\tnote{a}	&	-330	&	-8	&	-784	&	8	&	div	&	div\\
ELCC\tnote{b}	&	52	&	-47	&	101	&	-45	&	210	&	13\\
EACPF 	&	65	&	-19	&	123	&	-25	&	306	&	5\\
EAQCC 	&	68	&	-2	&	128	&	-13	&	347	&	1\\
EAQCCh	&	67	&	-8	&	127	&	-17	&	335	&	2\\	
\\
(Relaxed)\\
ECISD 	&	53	&	25	&	86	&	12	&	315	&	-8\\
ELCC\tnote{a}	&	-225	&	15	&	-582	&	-19	&	div	&	div\\
ELCC\tnote{b} 	&	-42	&	-8	&	-82	&	-18	&	-49	&	-6\\
EACPF 	&	-2	&	1	&	-6	&	-8	&	110	&	-7\\
EAQCC 	& 	17	&	7	&	29	&	-3	&	183	&	-7\\
EAQCCh 	& 	11	&	5	&	19	&	-4	&	161	&	-7\\
UCCSD	&	8	&	7	&	-15	&	14	&	67	&	-7\\
\\
Exptl.	&	1033	&	-19	&	1573&	19	&	588	&	-13\\
\hline\hline
\end{tabular} 
 {\footnotesize
\begin{tablenotes}
\item[a] Original ELCC with Eq.(\ref{eq:Q}).
\item[b] Corrected ELCC with Eq.(\ref{eq:generalQ}).
\end{tablenotes}
}
\end{threeparttable}
\end{table}

\subsection{Hyperfine coupling constants for BO, CO$^+$, and CN}\label{sec:HFCC}
The availability of relaxed density enables the computation of molecular properties with the ECEPA methods. In this section, we compute the isotropic hyperfine coupling constants (HFCC) of simple doublet molecules, BO, CO$^+$, and CN. Although these molecules are by no means strongly correlated (and therefore a spin-projection may not be as relevant), we aim to show that the size-consistent correction and relaxation effect are of critical importance for calculating properties in ECEPA. We also deem this section to serve as a good opportunity to demonstrate the significant consequences of treating open-shell orbitals improperly in ELCC.

The basis set used is EPR-III ([$7s4p2d$] for B and [$8s5p2d1f$] for other atoms), with the following geometries: $R_{\rm BO} = 1.2049 {\rm \AA}$, $R_{\rm CO} = 1.1287 {\rm \AA}$, and $R_{\rm CN} = 1.1718 {\rm \AA}$. No orbitals are frozen, because core correlation effects are important for HFCCs.

Table \ref{tb:HFCC} presents the errors from experimental HFCC values in MHz. It is apparent that dynamical correlation is indispensable to reproduce the experimental HFCC values. In this regard, SUHF improves the UHF results by adding a slight correlation and retaining qualitatively correct (symmetry-adapted) spin density, which is broken in UHF.
Interestingly, however, this is not always the case, and introducing spin-projection can sometimes deteriorate results; the HFCC of BO ($^{17}$O) increases significantly from 22 to 114 MHz with the wrong sign (the experimental value is -19 MHz). Note that this behavior is largely corrected when dynamical correlation is added in ECISD, but ECISD still gives the incorrect sign both with unrelaxed (24 MHz) and relaxed (6 MHz) densities.

As ECISD ameliorates the errors in SUHF, the size-consistent correction is considered important for molecular properties.
There is a solid improvement in EACPF and EAQCC over ECISD for both the unrelaxed and relaxed calculations.
As for ELCC, we consider two types of calculations.
In the first type, internal excitations are not properly treated; in other words, Eq.~(\ref{eq:Q}) is used for $\hat {Q}$, which is actually consistent with both SR- and MR-LCC methods.
However, it has come to our attention that this treatment gives rise to severe divergence in the calculations, and produces unphysical results, as is evident from Table \ref{tb:HFCC}.
We should emphasize that it is not the spin-projection that causes such an ill behavior---the traditional linearized coupled-cluster singles and doubles model poses exactly the same, or even worse, issues. Hence, in the second type of calculations, we adopt the same treatment for open-shell orbitals as in EACPF and EAQCC.
That is, Eq.~(\ref{eq:generalQ}) is used instead of Eq.~(\ref{eq:Q}).
ELCC corrected in this way provides more reasonable results for HFCCs (Table {\ref{tb:HFCC}}).
Nonetheless, the corrected ELCC does not outperform EACPF and EAQCC, as already pointed out in Sections \ref{sec:dissociation} and \ref{sec:ST}. EAQCCh always gives intermediate results between EACPF and EAQCC, as expected from its functional form in Eq.~(\ref{eq:AQCCh}).

We should also mention that the use of relaxed density generally shifts the unrelaxed results towards the experimental values for all methods.
For these ``well-behaved''  molecules with less static correlation, we find that EACPF performs better than EAQCC, although both are considered satisfactory.
In these situations, the EPV crudeness of EACPF is not as severe as in the dissociation cases, whereas the accuracy of dynamical correlation is slightly better than with AQCC averaging. Overall, EACPF predicts very similar HFCCs to those of UCCSD, which is encouraging.

\subsection{Cr$_2$}\label{sec:Cr2}
Finally, we discuss the results of ECEPA applied to the notorious chromium dimer.
The ground state ${^1}\Sigma_g^+$ of Cr$_2$ has posed a significant challenge in electronic structure theory, as it requires a very accurate description of both dynamical and static correlation effects. For this reason, a number of studies have attempted to elucidate this system using different methodologies.\cite{Andersson94,Roos95,Stoll96,Dachsel99,Muller09,Kurashige11,Coe14,Purwanto15,Carlson15,Vancoillie16, Guo16,Sokolov16} 

 We used 18 core orbitals for SUHF (all of which are almost doubly occupied, even without constraints) and then correlated the $3p$, $3d$, and $4s$ electrons in the subsequent ECISD/ECEPA calculations, thus utilizing 12 frozen-core orbitals. Our calculations were carried out with the cc-pV$n$Z basis sets with $n=$ D, T, and Q, and were extrapolated to the CBS limit using the two-point extrapolation formula.\cite{Helgaker97, Halkier98}
 Note that we did not correct the basis set superposition error (BSSE), which appears small enough in the CBS limit. Furthermore, no relativistic effects were included in our calculations, although it has been reported that the relativistic contribution at the equilibrium bond distance is non-negligible and  increases the dissociation energy by about {0.18--0.19~eV} in both CASPT2\cite{Andersson94,Roos95} and MR-ACPF.\cite{ Stoll96} In any case, it is not our intention to obtain accurate dissociation energies; we are more interested in the qualitative behaviors of the proposed methods.

\begin{figure}[t!]
\includegraphics[width=85mm, bb=0 0 277 384]{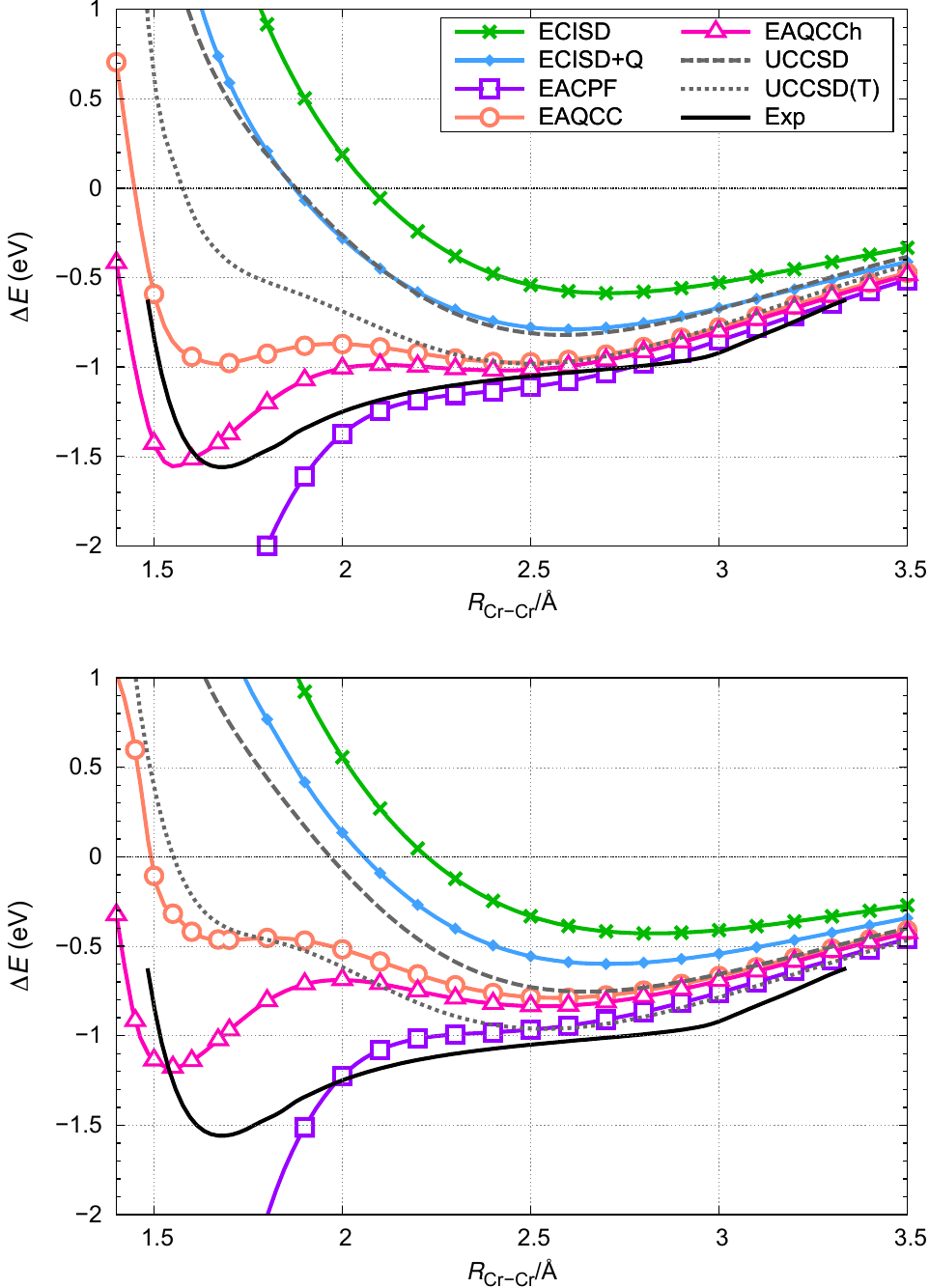}
\caption{Potential curves of Cr$_2$ with cc-pVDZ (top) and at the CBS limit (bottom).}\label{fig:Cr2}
\end{figure}

In the top panel of Figure \ref{fig:Cr2}, we depict the potential curves of Cr$_2$ in cc-pVDZ at several levels of theory. As is well known, UCCSD and UCCSD(T) fail to predict the inner minimum (experimental value is $R_e = 1.679$~${\rm \AA}$\cite{Cr2Exp}). The results of ECISD and ECISD+Q are disappointing, with almost no hint of the true minimum, similar to UCCSD. However, it is interesting to see that both EACPF and EAQCC show completely different potential curves. We again observe the same behaviors in the two methods, which are now more noticeable than ever. 
EACPF substantially overcorrelates around $R_e$ because of the inadequate treatment of the EPV contribution.
EAQCC, in contrast, is missing some dynamical correlation, but captures the correct shape of the experimental curve. However, with a small basis set such as cc-pVDZ, BSSE is rather large; the TZ/QZ extrapolation to CBS predicts a shallower curve with EAQCC (the bottom panel of Figure \ref{fig:Cr2}), although it is still bound at the inner minimum. EAQCCh again provides a fine compromise, yielding a qualitatively correct curve. Given the relativistic contribution of $\sim$0.2 eV, the dissociation energy of EAQCCh would appear somewhere around 1.4 eV, which is consistent with the experimental value of 1.45 $\sim$ 1.57 eV.\cite{Cr2Exp1,Cr2Exp2,Cr2Exp3}
However, the estimate of the bond-length (ca. 1.55 $\rm \AA$) is too short compared with the experimental value and EAQCC's inner minimum (1.65$\sim$1.7 $\rm\AA$). 

Nevertheless, we argue that these results are still encouraging, because ECEPA is able to achieve a qualitative description of the difficult case of Cr$_2$ where both ECISD and ECISD+Q completely fail. The reference state, SUHF, is very simple, as it can be considered as a linear combination of {\it four} nonorthogonal Slater determinants (recall that we only use $N_{\rm grid} = 4$ for the numerical integration of $\hat {\cal P}$). This is a considerable simplification from the minimum CAS reference, which contains hundreds of thousands of configurations.
However, at the same time, we realize the limitation of the average ECEPA schemes; to obtain more accurate descriptions, one will need to abandon such averaging, or treat the quadratic (and possibly higher) terms explicitly without approximations. Another possibility is to use a linear combination of a few SUHF states, which no doubt promises a better reference.\cite{Rodriguez13,Jimenez13B} Either approach is expected to give an improved description of Cr$_2$.

\section{Discussion}\label{sec:Discussions}
\subsection{Taylor expansion of energy functional}\label{sec:Taylor}
Despite its {\it a posteriori} treatment, it was shown that in some cases ECISD+Q can perform better than the ECEPA methods (Section \ref{sec:dissociation}). Here, we show that this is simply a fortuitous error cancellation.

The Davidson correction $\Delta E_{\rm Q}$ to the ``variational'' ECISD correlation energy $E_c^{\rm var}$ (with respect to the SUHF energy $E_{\rm SUHF}$) is  given by
\begin{align}
\Delta E_{\rm Q} &= \left(1- \bra \Phi_0 | \hat {\cal P}  | \Psi^{\rm ECISD}\ket^2\right) E_c^{\rm var}[\Psi^{\rm ECISD}] \bre
&= \bra \Psi^{\rm ECISD} | \hat {\cal P} \hat {Q} | \Psi^{\rm ECISD}\ket  E_c^{\rm var}[\Psi^{\rm ECISD}],
\end{align}
with 
\begin{align}
E_c^{\rm var}[\Psi] =\bra  \Psi | \hat {\bar H} \hat {\cal P} | \Psi \ket,
\end{align}
where both $\bra \Phi_0|\hat {\cal P}|\Phi_0\ket$ and $\bra  \Psi^{\rm ECISD}  |\hat {\cal P} | \Psi^{\rm ECISD} \ket$ are properly normalized.
The Taylor expansion of the ELCC correlation energy $F_c^{\rm ELCC} = F[\zeta=1] - E_{\rm SUHF}$ around $E_c^{\rm var}[\tilde \Psi]$, provided that $\bra \tilde \Psi |\hat {\cal P} \hat {Q}|\tilde\Psi\ket \ll 1$, is: 
\begin{align}
 F_c^{\rm ELCC} 
 &= \frac{\bra \tilde \Psi | \hat {\bar H} \hat {\cal P} |\tilde \Psi\ket  }{1-  \bra \tilde \Psi|\hat{\cal P}\hat{Q}|\tilde \Psi\ket }\bre
&= \bra \tilde \Psi | \hat {\bar H} \hat {\cal P} |\tilde \Psi\ket \left(1+ \bra \tilde \Psi|\hat{\cal P}\hat{Q}|\tilde \Psi\ket +  \bra \tilde \Psi|\hat{\cal P}\hat{Q}|\tilde \Psi\ket^2 + \cdots \right)\bre
&= E_c^{\rm var}[\tilde\Psi]  + \Delta E_{\rm Q}[\tilde \Psi] +  {\cal O}\left( \bra \tilde \Psi|\hat{\cal P}\hat{Q}|\tilde \Psi\ket^2\right).\label{eq:FELCC}
\end{align}  
Therefore, the correlation energy of ECISD+Q appears in the first-order contribution of $F_c^{\rm ELCC}$.
We should mention that the ELCC energy is always lower than that of ECISD+Q, because (1) the higher-order terms in Eq.~(\ref{eq:FELCC}) are necessarily negative and (2) $\tilde\Psi$ is variationally determined by minimizing $F_c^{\rm ELCC}$, whereas ECISD+Q uses the wave function obtained {\it a priori} according to $E_c^{\rm var}$. We have seen in the previous sections that Eq.~(\ref{eq:FELCC}) is predisposed to significant overestimation when variationally optimized, and hence terminating $F_c^{\rm ELCC}$ at the first order in ECISD+Q and discarding the higher-order terms seems to be a reasonable compromise. 

It is also worth pointing out that similar relations can be found for EACPF and EAQCC. By expanding $F_c^{\rm EACPF}$ and $F_c^{\rm EAQCC}$, the first-order terms turn out to correspond to the Pople-\cite{Pople77} and Meissner-type\cite{Meissner88,Szalay93} corrections to ECISD.\cite{Tsuchimochi16B}

\subsection{Size-consistency and size-extensivity}\label{sec:SC}
In this subsection, we study the size-consistent and size-extensive properties of ECEPA and ECISD using some numerical evidence.
Let us first recall the general definitions of size-consistency and size-extensivity. Size-consistency is a property of a method whereby the energy calculated for a supersystem composed of sufficiently separated local subsystems $X$ and $Y$ is the sum of the energies computed independently for each subsystem, i.e., $E^{X+Y} = E^X + E^Y$. Size-extensivity is a somewhat weaker requirement, defined as a mathematical property that guarantees the proper scaling behavior of extensive physical quantities such as energy with the number of electrons. Size-consistency is a very difficult requirement to achieve, especially if $X$ and $Y$ are quantum-mechanically entangled, as seen in the dissociation limit of chemical bonds. Size-extensivity in the correlation energy is one of the most important prerequisites for a post-mean-field method to be applicable to extended systems. Do our methods fulfill these requirements?

Let us consider the case where the underlying broken-symmetry wave function  $| \Psi \ket$ is separable into local subsystems $X$ and $Y$. This means
\begin{align}
 |\Psi\ket =  |\Psi^X\ket  |\Psi^Y\ket. \label{eq:XY}
\end{align} 
If a projection operator is also separable, one can write
\begin{align}
\hat {\cal P} |\Psi \ket =  \hat {\cal P}^{X+Y} |\Psi\ket =  \hat {\cal P}^X |\Psi^X\ket \hat {\cal P}^Y |\Psi^Y\ket,
\end{align}
and its intermediate norm with a Slater determinant $|\Phi_0\ket$ becomes a product of local quantities:  
\begin{align}
\bra \Phi^X| \bra \Phi^Y | \hat {\cal P}^{X+Y} |  \Psi^X\ket | \Psi^Y\ket &= \bra \Phi^X | \hat {\cal P}^X |  \Psi^X\ket  \bra \Phi^Y | \hat {\cal P}^{Y}|  \Psi^Y \ket.
\end{align}
Given that $\hat H = \hat H^X + \hat H^Y$, the projective correlation energy Eq.~(\ref{eq:projectiveEc}) is
\begin{align}
E_c^{X+Y}&= \frac{\bra \Phi^X| \bra \Phi^Y| \bar  H \hat {\cal P}|  \Psi^X \ket | \Psi^Y\ket}{\bra \Phi^X| \bra \Phi^Y | \hat {\cal P}|  \Psi^X \ket|  \Psi^Y\ket}\bre
&= E_c^X +  E_c^Y,\label{eq:SC}
\end{align}
which ensures size-consistency in the correlation energy. Note that if the separability in either $|\Psi\ket$ or $\hat{\cal P}$ is not fulfilled, size-consistency is not guaranteed. As the latter is not separable in general, the $X$ and $Y$ components couple with each other through both the Hamiltonian and the overlap. However, if ${\hat {\cal P}}$ acts only on the $Y$ component or {\it if $| \Psi^X\ket$ is already symmetry-adapted}, one can write $\hat {\cal P} |\Psi^X\ket |\Psi^Y\ket \equiv |\Psi^X\ket \otimes \hat {\cal P} |\Psi^Y\ket$, and the equality in Eq.~(\ref{eq:SC}) evidently holds. In such a case, any projected wave function satisfying Eq.~(\ref{eq:XY}) is size-consistent. ECISD obviously does not hold this property, and is therefore expected  to suffer from a large size-inconsistency error, $\Delta_{\rm SC} = E_c^{X+Y} - \left(E_c^{X}+E_c^{Y}\right)$. Here, we are interested in how the size-inconsistency error in ECISD can be reduced by adopting the cluster approximation in ECEPA.

\begin{table*}
\caption{Size-inconsistency error $\Delta_{\rm SC}$ of Be$_2$ for each supersystem (m$E_h$).}\label{tb:SC}
\begin{tabular}{cccccccccc}
\hline\hline
$X$ & $Y$ & SUHF & ECISD & ECISD+Q & ELCC & EACPF & EAQCC\\
\hline
A &A & 0.00&	7.84&	3.92&	0.00&	0.00&	0.93\\
A& B & 0.00&	5.18&	2.27&	0.00&	-0.22&	0.32\\
B& B & 8.26 &	3.52&	-1.64	&	-5.55&	-3.36&	-1.66\\
\hline\hline
\end{tabular}
\end{table*}

To test the size-consistency in each method, we carried out very simple calculations with the infinitely separated Be dimer using 6-31G. Note that there is no entanglement between each atom, and thus the system is non-interacting. Below, we consider three different supersystems in which the symmetries of subsystems $X$ and $Y$ are either adapted (A) or broken and then projected (B); $XY = {\rm AA}, {\rm AB}$, or ${\rm BB}$. Note that all the methods considered correspond to their single-reference limit for the AA model. 

We summarize the size-inconsistency errors $\Delta_{\rm SC}$ of each method applied to these model systems in Table \ref{tb:SC}. As a Slater determinant is written as Eq.~(\ref{eq:XY}), SUHF is size-consistent for AA and AB. However, it  acquires a large size-inconsistency error for BB of 8.26 m$E_h$, as a consequence of the non-separability of $\hat {\cal P}$. ECISD is a truncated CI, which is known to be size-inconsistent.\cite{Duch94}  Hence, this gives a large $\Delta_{\rm SC}$ of 5.18 m$E_h$ for AB, which nonetheless is slightly smaller than that of conventional CISD for AA (7.84 m$E_h$). Furthermore, it is worth mentioning that, for BB, the error in SUHF is compensated by ECISD. This is very much consistent with the observation made in our previous study\cite{Tsuchimochi16A}; although size-inconsistent itself, ECISD indeed mitigates some of the size-inconsistency effects in SUHF. The error in ECISD can be further reduced by introducing the {\it a posteriori} size-consistent correction in ECISD+Q for all cases. 

In contrast, the simplest ECEPA, ELCC, successfully cancels out $\Delta_{\rm SC}$ in AA and AB by effectively forming a cluster wave function, but over-corrects the error in BB. There are two competing error sources: spin-projection and overemphasis of higher excitation effects. Although it is hard to identify and quantify the relative contributions of these two, the latter seems to make the errors more negative by overestimating the correlation energy. Indeed, methods with more sophisticated EPV terms (EACPF and EAQCC) produce reasonably small errors in BB compared with that given by ELCC, albeit at the cost of losing precise size-consistency for the other cases. Overall, we find EAQCC is the best compromise, as it is approximately size-consistent for all cases. 

\begin{figure}
\includegraphics[width=85mm, bb=0 0 350 245]{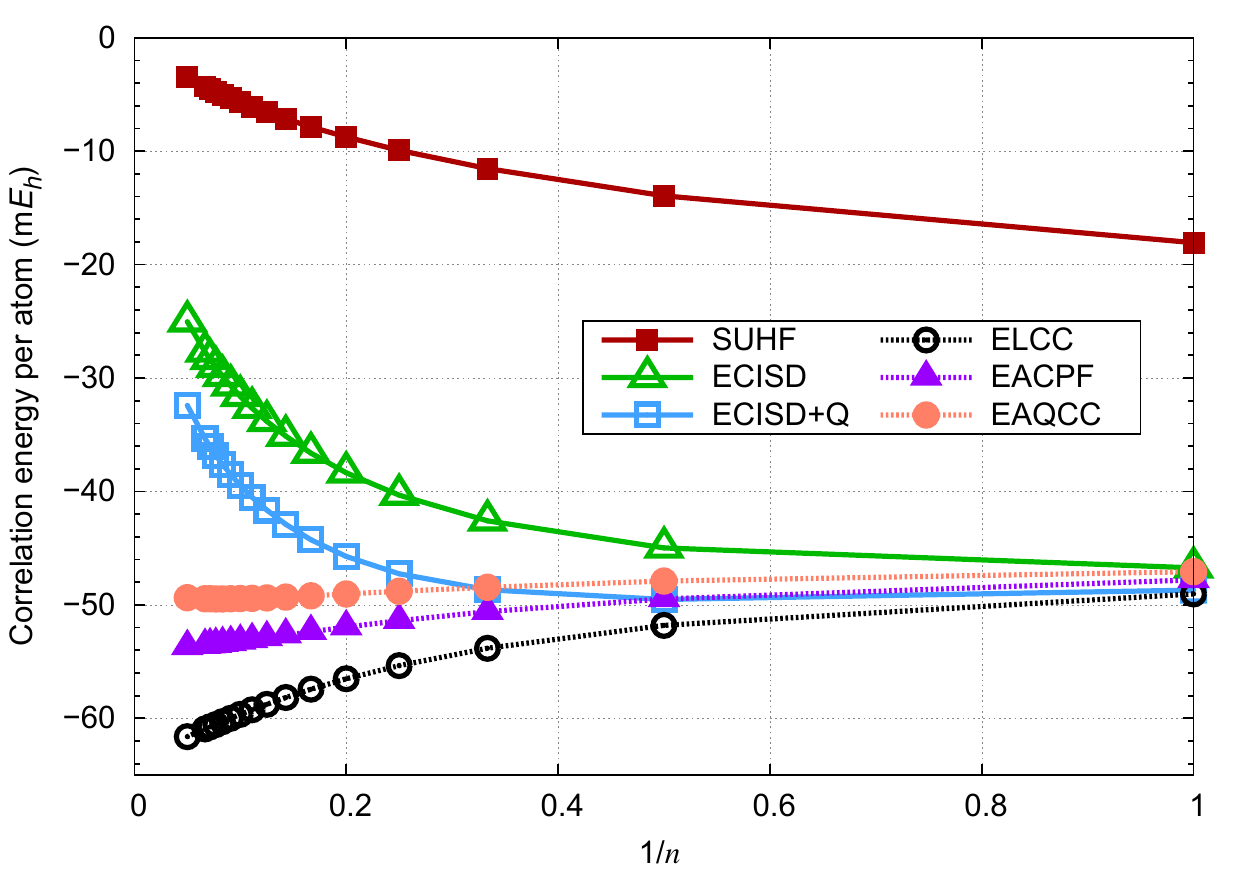}
\caption{Correlation energy per atom for non-interacting Be$_n$. Spin-symmetry in each atom is broken and restored.}\label{fig:Be_c}
\end{figure}

The size-extensivity of each method can be studied using $n$ non-interacting Be atoms ($n>1$),\cite{Duch94}  where all the atoms experience broken-symmetry. The correlation energy {\it per atom} (with respect to HF) is plotted for each method in Figure \ref{fig:Be_c}, in which we set $N_{\rm grid} = 7$ to perform precise spin-projection. Because of the lack of size-extensivity, it is expected that SUHF, ECISD, and ECISD+Q will all asymptotically converge to the zero correlation limit as $n\rightarrow\infty$. However, ECEPA methods clearly exhibit different behavior, namely, they preserve an approximately constant correlation energy per atom, manifesting the size-extensive property correctly captured in ECEPA. We find that ELCC might be divergent, and  EAQCC is most stable.

Finally, we should mention that, if and only if the high-spin state is set to triplet for all $n$, Yamaguchi's approximate spin-projection will yield exactly size-consistent and size-extensive results for these simple systems via fortuitous error cancellations. However, when the subsystems are not identical, the size-consistency is generally lost in the BB system. For example, if subsystems $X$ and $Y$ both correspond to H$_2$ molecules but with different separations, $R_X =2.0$ ${\rm \AA}$ and $ R_Y = 1.5$ ${\rm \AA}$, the approximation gives rise to a large size-inconsistency error, to a similar degree of SUHF.

\section{Concluding remarks}\label{sec:Conclusions}
The recently developed spin-projected extension of CI bridges the gap between SR-CI and MR-CI, enabling the accurate reproduction of the latter, especially when the Davidson correction is incorporated.
However, the method has a fundamental obstacle that it is neither size-consistent nor size-extensive, and therefore cannot be used as a predictive tool for a wide range of chemical systems. 

In this respect, the present work marks an important step forward, introducing the size-consistency and size-extensivity effects by correcting the unlinked terms in the ECISD equation. Such a reformulation was successfully achieved via cluster excitations, without explicitly working out the higher excitation terms.
By analogy with both single- and multireference developments, several variants of ECEPA were proposed, depending on the approximation used for treating the EPV terms.  In this paper, we focused on the averaged ECEPA schemes because they have a well-defined energy functional that is unitary invariant with orbital rotation within the occupied space. The variational nature of the ECEPA energy functionals fulfills the generalized Hellmann--Feynman theorem, and thus enables straightforward evaluation of the energy gradient to compute the relaxed density matrix.

The proposed methods were tested with several bond-dissociation curves, singlet-triplet splitting gaps, and hyperfine coupling constants. ECEPA showed remarkable improvements over ECISD when the latter is unacceptable, and otherwise retained the same accuracy as ECISD. Furthermore, the application to the Cr$_2$ potential energy curve revealed that an appropriate treatment of higher excitation effects is essential for describing the correct shape in spin-extended methods. ECISD and ECISD+Q both strongly underbound the molecule, similar to UCCSD. Although dynamical correlation is overemphasized in EACPF and underestimated in EAQCC, the newly proposed averaging scheme, EAQCCh, was found to be a good compromise, giving a reasonable potential curve for Cr$_2$. However, the large difference between EACPF and EAQCC in Cr$_2$ demonstrates the limitation of these averaged treatments of the EPV terms, as well as the possible inadequacy of SUHF as a reference. 
Hence, although ECEPA is (approximately) size-consistent and size-extensive, the method needs further refinement. Possible research directions include more appropriate treatments of cluster excitations and the use of an improved reference such as a linear combination of SUHF wave functions.

\section*{Acknowledgements}
This work was supported by MEXT's FLAGSHIP2020 as priority issue 5 (development of new fundamental technologies for high-efficiency energy creation, conversion/storage and use). We are grateful to the HPCI System Research project for the use of computer resources to carry out some of the calculations (Project ID: hp150228, hp150278).

\bibliographystyle{aip}

\begin{thebibliography}{10}

\bibitem{Cizek66}
J.~\v{C}\'i\v{z}ek,
\newblock J. Chem. Phys. {\bf 45}, 4256 (1966).

\bibitem{Bartlett07}
R.~J. Bartlett and M.~Musia\l,
\newblock Rev. Mod. Phys. {\bf 79}, 291 (2007).

\bibitem{Shavitt09}
I.~Shavitt and R.~J. Bartlett,
\newblock {\em Many-Body Methods in Chemistry and Physics: MBPT and
  Coupled-Cluster Theory},
\newblock Cambridge Universiy Press, Cambridge, UK, 2009.

\bibitem{Bulik15}
I.~W. Bulik, T.~M. Henderson, and G.~E. Scuseria,
\newblock J. Chem. Theory Comput. {\bf 11}, 3171 (2015).

\bibitem{Degroote16}
M.~Degroote, T.~M. Henderson, J.~Zhao, J.~Dukelsky, and G.~E. Scuseria,
\newblock Phys. Rev. B {\bf 93}, 125124 (2016).

\bibitem{Roos80}
B.~O. Roos, P.~R. Taylor, and P.~E. Siegbahn,
\newblock Chem. Phys. {\bf 48}, 157 (1980).

\bibitem{Siegbahn80}
P.~Siegbahn, A.~Heiberg, B.~Roos, and B.~Levy,
\newblock Physica Scripta {\bf 21}, 323 (1980).

\bibitem{Lyakh11}
D.~I. Lyakh, M.~Musia\l, V.~F. Lotrich, and R.~J. Bartlett,
\newblock Chem. Rev. {\bf 112}, 182 (2012).

\bibitem{Kohn12}
A.~K\"ohn, M.~Hanauer, L.~A. M\"uck, T.-C. Jagau, and J.~Gauss,
\newblock WIREs Comput. Mol. Sci. {\bf 3}, 176.

\bibitem{Scuseria11}
G.~E. Scuseria, C.~A. Jim\'enez-Hoyoz, T.~M. Henderson, K.~Samanta, and J.~K.
  Ellis,
\newblock J. Chem. Phys. {\bf 135}, 124108 (2011).

\bibitem{Jimenez12}
C.~A. Jim$\rm\acute{e}$nez-Hoyoz, T.~M. Henderson, T.~Tsuchimochi, and G.~E.
  Scuseria,
\newblock J. Chem. Phys. {\bf 136}, 164109 (2012).

\bibitem{Tsuchimochi14}
T.~Tsuchimochi and T.~Van~Voorhis,
\newblock J. Chem. Phys. {\bf 141}, 164117 (2014).

\bibitem{Tsuchimochi15A}
T.~Tsuchimochi and T.~Van~Voorhis,
\newblock J. Chem. Phys. {\bf 142}, 124103 (2015).

\bibitem{Tsuchimochi15C}
T.~Tsuchimochi,
\newblock J. Chem. Phys. {\bf 143}, 144114 (2015).

\bibitem{Tsuchimochi16A}
T.~Tsuchimochi and S.~Ten-no,
\newblock J. Chem. Phys. {\bf 144}, 011101 (2016).

\bibitem{Tsuchimochi16B}
T.~Tsuchimochi and S.~Ten-no,
\newblock J. Chem. Theory Comput. {\bf 12}, 1741 (2016).

\bibitem{Lowdin55B}
P.-O. L$\rm\ddot{o}$wdin,
\newblock Phys. Rev. {\bf 97}, 1509 (1955).

\bibitem{Mayer73}
I.~Mayer, J.~Ladik, and G.~Bicz$\rm\acute{o}$,
\newblock Int. J. Quantum Chem. {\bf 7}, 583 (1973).

\bibitem{Mayer80}
I.~Mayer,
\newblock Adv. Quantum Chem. {\bf 12}, 189 (1980).

\bibitem{Bartlett89}
R.~J. Bartlett,
\newblock J. Phys. Chem. {\bf 93}, 1697 (1989).

\bibitem{Davidson74}
E.~R. Davidson,
\newblock {\em The World of Quantum Chemistry},
\newblock Reidel, Dordrecht, 1974.

\bibitem{Langhoff74}
S.~R. Langhoff and E.~R. Davidson,
\newblock Int. J. Quantum Chem. {\bf 8}, 61 (1974).

\bibitem{Pople77}
I.~A. Pople, J.~S. Binkley, and R.~Seeger,
\newblock Int. J. Quantum Chem. {\bf S11}, 149 (1977).

\bibitem{Meissner88}
L.~Meissner,
\newblock Chem. Phys. Lett. {\bf 146}, 204 (1988).

\bibitem{Duch94}
W.~Duch and G.~H.~F. Diercksen,
\newblock J. Chem. Phys. {\bf 101}, 3018 (1994).

\bibitem{Tsuchimochi17A}
T.~Tsuchimochi and S.~Ten-no,
\newblock arXiv:1612.03521.

\bibitem{Meyer71}
W.~Meyer,
\newblock Int. J. Quantum Chem. {\bf S5}, 341 (1971).

\bibitem{Meyer73}
W.~Meyer,
\newblock J. Chem. Phys. {\bf 58}, 1017 (1973).

\bibitem{Ahlrichs79}
R.~Ahlrichs,
\newblock Comput. Phys. Commun. {\bf 17}, 31 (1979).

\bibitem{Ahlrichs85}
R.~Ahlrichs, P.~Scharf, and C.~Ehrhardt,
\newblock J. Chem. Phys. {\bf 82}, 890 (1985).

\bibitem{Ahlrichs87}
R.~Ahlrichs and P.~Scharf,
\newblock in {\em Ab Initio Methods in Quantum Chemistry}, edited by K.~P.
  Lawley, page 501, Wiley, 1987.

\bibitem{Daudey93}
J.~Daudey, J.~Heully, and J.~Malrieu,
\newblock J. Chem. Phys. {\bf 99}, 1240 (1993).

\bibitem{Wennmohs08}
F.~Wennmohs and F.~Neese,
\newblock Chem. Phys. {\bf 343}, 217 (2008).

\bibitem{Kallay00}
M.~Kallay and P.~R. Surjan,
\newblock J. Chem. Phys. {\bf 113}, 1359 (2000).

\bibitem{Hirata00}
S.~Hirata and R.~J. Bartlett,
\newblock Chem. Phys. Lett. {\bf 321}, 216 (2000).

\bibitem{Hohenberg64}
P.~Hohenberg and W.~Kohn,
\newblock Phys. Rev. {\bf 136}, B864 (1964).

\bibitem{Kohn65}
W.~Kohn and L.~J. Sham,
\newblock Phys. Rev. {\bf 140}, A1133 (1965).

\bibitem{Noodleman81}
L.~Noodleman,
\newblock J. Chem. Phys. {\bf 74}, 5737 (1981).

\bibitem{Noodleman86}
L.~Noodleman and E.~R. Davidson,
\newblock Chem. Phys. {\bf 109}, 131 (1986).

\bibitem{Yamaguchi86}
K.~Yamaguchi, Y.~Takahara, and T.~Fueno,
\newblock Ab-initio molecular orbital studies of structure and reactivity of
  transition metal-oxo compounds,
\newblock in {\em Applied Quantum Chemistry}, edited by V.~H. {Smith Jr.},
  H.~F. {Schaefer III}, and K.~Morokuma, pages 155--184, Springer Netherlands,
  1986.

\bibitem{Yamaguchi88}
K.~Yamaguchi, F.~Jensen, A.~Dorigo, and K.~N. Houk,
\newblock Chem. Phys. Lett. {\bf 149}, 537 (1988).

\bibitem{Yamanaka94}
S.~Yamanaka, T.~Kawakami, H.~Nagao, and K.~Yamaguchi,
\newblock Chem. Phys. Lett. {\bf 231}, 25 (1994).

\bibitem{Nishino97}
M.~Nishino, S.~Yamanaka, Y.~Yoshioka, and K.~Yamaguchi,
\newblock J. Phys. Chem. A {\bf 101}, 705 (1997).

\bibitem{Saito10}
T.~Saito et~al.,
\newblock J. Phys. Chem. A {\bf 114}, 7967 (2010).

\bibitem{Szabo}
A.~Szabo and N.~S. Ostlund,
\newblock {\em Modern Quantum Chemistry: Introduction to Advanced Electronic
  Structure Theory},
\newblock Dover Publications, Mineola, NY, 1996.

\bibitem{Koch81}
S.~Koch and W.~Kutzelnigg,
\newblock Theor. Chim. Acta {\bf 59}, 387 (1981).

\bibitem{Bartlett1981}
R.~J. Bartlett,
\newblock Annu. Rev. Phys. Chem. {\bf 32}, 359 (1981).

\bibitem{Laidig84}
W.~D. Laidig and R.~J. Bartlett,
\newblock Chem. Phys. Lett. {\bf 104}, 424 (1984).

\bibitem{Laidig87}
W.~D. Laidig, P.~Saxe, and R.~J. Bartlett,
\newblock J. Chem. Phys. {\bf 86}, 88=7 (1987).

\bibitem{Taube09}
A.~G. Taube and R.~J. Bartlett,
\newblock J. Chem. Phys. {\bf 130}, 144112 (2009).

\bibitem{Kelly63}
H.~P. Kelly and A.~M. Sessler,
\newblock Phys. Rev. {\bf 132}, 2091 (1963).

\bibitem{Kelly64}
H.~P. Kelly,
\newblock Phys. Rev. {\bf 134}, A1450 (1964).

\bibitem{Meyer74}
W.~Meyer,
\newblock Theor. Chim. Acta {\bf 35}, 277 (1974).

\bibitem{Gdanitz87}
R.~J. Gdanitz and R.~Ahlrichs,
\newblock Chem. Phys. Lett. {\bf 143}, 413 (1988).

\bibitem{Szalay93}
P.~G. Szalay and R.~J. Bartlett,
\newblock Chem. Phys. Lett. {\bf 214}, 481 (1993).

\bibitem{Fusti96}
L.~F\"usti-Moln\'ar and P.~G. Szalay,
\newblock J. Phys. Chem. A {\bf 100}, 6288 (1996).

\bibitem{Szalay95}
P.~G. Szalay and R.~J. Bartlett,
\newblock J. Chem. Phys. {\bf 103}, 3600 (1995).

\bibitem{Gdanitz01}
R.~J. Gdanitz,
\newblock Int. J. Quantum Chem. {\bf 85}, 281 (2001).

\bibitem{Ruttink91}
P.~J.~A. Ruttink, J.~H. {van Lenthe}, R.~Zwaans, and G.~C. Groenenboom,
\newblock J. Chem. Phys. {\bf 94}, 7212 (1991).

\bibitem{Malrieu94}
J.~Malrieu, J.~Daudey, and R.~Caballol,
\newblock J. Chem. Phys. {\bf 101}, 8908 (1994).

\bibitem{Tsuchimochi10B}
T.~Tsuchimochi and G.~E. Scuseria,
\newblock J. Chem. Phys. {\bf 133}, 141102 (2010).

\bibitem{Tsuchimochi11}
T.~Tsuchimochi and G.~E. Scuseria,
\newblock J. Chem. Phys. {\bf 134}, 064101 (2011).

\bibitem{Heully92}
J.-L. Heully and J.-P. Malrieu,
\newblock Chem. Phys. Lett. {\bf 199}, 545 (1992).

\bibitem{Helgaker89}
T.~Helgaker and P.~J{\o}rgensen,
\newblock Theor. Chim. Acta {\bf 75}, 111 (1989).

\bibitem{Jorgensen88}
P.~J{\o}rgensen and T.~Helgaker,
\newblock J. Chem. Phys. {\bf 89}, 1560 (1988).

\bibitem{Szalay00}
P.~G. Szalay, T.~M\"uller, and H.~Lischka,
\newblock Phys. Chem. Chem. Phys. {\bf 2}, 2067 (2000).

\bibitem{JJSakurai}
J.~J. Sakurai,
\newblock {\em Modern Quantum Mechanics},
\newblock Addison-Wesley, Massachusetts, 1994.

\bibitem{RingSchuck}
P.~Ring and P.~Schuck,
\newblock {\em The Nuclear Many-Body Problem},
\newblock Springer-Verlag, Berlin, 1980.

\bibitem{Tsuchimochi10A}
T.~Tsuchimochi, T.~M. Henderson, G.~E. Scuseria, and A.~Savin,
\newblock J. Chem. Phys. {\bf 133}, 134108 (2010).

\bibitem{Bao16}
J.~L. Bao, A.~Sand, L.~Gagliardi, and D.~G. Truhlar,
\newblock J. Chem. Theory Comput. {\bf 12}, 4274 (2016).

\bibitem{Andersson90}
K.~Andersson, P.-{\rm\AA}. Malmqvist, B.~O. Roos, A.~J. Sadlej, and
  K.~Wolinski,
\newblock J. Phys. Chem. {\bf 94}, 5483 (1990).

\bibitem{Andersson92}
K.~Andersson, P.-{\rm\AA}. Malmqvist, and B.~O. Roos,
\newblock J. Chem. Phys. {\bf 96}, 1218 (1992).

\bibitem{Ghigo04}
G.~Ghigo, B.~O. Roos, and P.-{\rm\AA}. Malmqvist,
\newblock Chem. Phys. Lett. {\bf 396}, 142 (2004).

\bibitem{PBE}
J.~P. Perdew, K.~Burke, and M.~Ernzerhof,
\newblock Phys. Rev. Lett. {\bf 77}, 3865 (1996).

\bibitem{Manni14}
G.~L. Manni et~al.,
\newblock J. Chem. Theory Comput. {\bf 10}, 3669 (2014).

\bibitem{Carlson15}
R.~K. Carlson, D.~G. Truhlar, and L.~Gagliardi,
\newblock J. Chem. Theory Comput. {\bf 11}, 4077 (2015).

\bibitem{Chan15}
B.~Chan and L.~Radom,
\newblock J. Chem. Theory Comput. {\bf 11}, 2109 (2015).

\bibitem{Ko13}
K.~C. Ko, D.~Cho, and J.~Y. Lee,
\newblock J. Phys. Chem. A {\bf 117}, 3561 (2013).

\bibitem{Cho15}
D.~Cho et~al.,
\newblock J. Chem. Phys. {\bf 142}, 024318 (2015).

\bibitem{Andersson94}
K.~Andersson, B.~O. Roos, P.-{\rm\AA}. Malmqvist, and P.-O. Widmark,
\newblock Chem. Phys. Lett. {\bf 230}, 391 (1994).

\bibitem{Roos95}
B.~O. Roos and K.~Andersson,
\newblock Chem. Phys. Lett. {\bf 245}, 215 (1995).

\bibitem{Stoll96}
Stoll and Werner,
\newblock Mol. Phys. {\bf 88}, 793 (1996).

\bibitem{Dachsel99}
H.~Dachsel, R.~J. Harrison, and D.~A. Dixon,
\newblock J. Phys. Chem. A {\bf 103}, 152 (1999).

\bibitem{Muller09}
T.~M\"uller,
\newblock J. Phys. Chem. A {\bf 113}, 12729 (2009).

\bibitem{Kurashige11}
Y.~Kurashige and T.~Yanai,
\newblock J. Chem. Phys. {\bf 135}, 094104 (2011).

\bibitem{Coe14}
J.~Coe, P.~Murphy, and M.~Paterson,
\newblock Chem. Phys. Lett. {\bf 604}, 46 (2014).

\bibitem{Purwanto15}
W.~Purwanto, S.~Zhang, and H.~Krakauer,
\newblock J. Chem. Phys. {\bf 142}, 064302 (2015).

\bibitem{Vancoillie16}
S.~Vancoillie, P.~{\rm\AA}. Malmqvist, and V.~Veryazov,
\newblock J. Chem. Theory Comput. {\bf 12}, 1647 (2016).

\bibitem{Guo16}
S.~Guo et~al.,
\newblock J. Chem. Theory Comput. {\bf 12}, 1583 (2016).

\bibitem{Sokolov16}
A.~Y. Sokolov and G.~K.-L. Chan,
\newblock J. Chem. Phys. {\bf 144}, 064102 (2016).

\bibitem{Helgaker97}
T.~Helgaker, W.~Klopper, H.~Koch, and J.~Noga.

\bibitem{Halkier98}
A.~Halkier et~al.

\bibitem{Cr2Exp}
S.~M. Casey and D.~G. Leopold,
\newblock J. Phys. Chem. {\bf 97}, 816 (1993).

\bibitem{Cr2Exp1}
C.-X. Su, D.~A. Hales, and P.~Armentrout,
\newblock Chem. Phys. Lett. {\bf 201}, 199 (1993).

\bibitem{Cr2Exp2}
K.~Hilpert and R.~Ruthardt,
\newblock Ber. Bunsenges. Phys. Chem. {\bf 91}, 724 (1987).

\bibitem{Cr2Exp3}
B.~Simard, M.-A. Lebeault-Dorget, A.~Marijnissen, and J.~J. ter Meulen,
\newblock J. Chem. Phys. {\bf 108}, 9668 (1998).

\bibitem{Rodriguez13}
R.~Rodr$\rm\acute{i}$guez-Guzm$\rm\acute{a}$n, C.~A.
  Jim$\rm\acute{e}$nez-Hoyos, R.~Schutski, and G.~E. Scuseria,
\newblock Phys. Rev. B {\bf 87}, 235129 (2013).

\bibitem{Jimenez13B}
C.~A. Jim$\rm\acute{e}$nez-Hoyos,
  R.~Rodr$\rm\acute{i}$guez-Guzm$\rm\acute{a}$n, and G.~E. Scuseria,
\newblock J. Chem. Phys. {\bf 139}, 204102 (2013).

\end{thebibliography}

\end{document}